\documentclass{article}
\usepackage{arxiv}

\usepackage{amsmath,amsfonts,amssymb}
\usepackage{mathtools}
\usepackage[hidelinks]{hyperref}
\usepackage{algorithmic}
\usepackage{algorithm}
\usepackage{array}
\usepackage[caption=false,font=normalsize,labelfont=sf,textfont=sf]{subfig}
\usepackage{relsize}
\usepackage{textcomp}
\usepackage{stfloats}
\usepackage{url}
\usepackage{verbatim}
\usepackage{graphicx}
\usepackage{natbib}
\usepackage{doi}
\usepackage{xcolor}
\usepackage{paralist}
\usepackage{blindtext}
\usepackage{multicol}
\usepackage{multirow}
\usepackage[subnum]{cases}

\newtheorem{theorem}{Theorem}[section]

\newtheorem{remark}[theorem]{Remark}

\title{Traffic Smoothing Through Automated Vehicle Control With Optimal Parameter Selection}

\author{Shian Wang\\
	Civil, Environmental, and Architectural Engineering\\
	The University of Kansas\\
	\texttt{shian.wang@ku.edu} \\
	\And
	Jose Carlos Acedo Aguilar \\
	Electrical and Computer Engineering\\
	The University of Texas at El Paso\\
	\texttt{jcacedoagui@miners.utep.edu} \\
	\And
	Miguel Velez-Reyes \\
	Electrical and Computer Engineering\\
	The University of Texas at El Paso\\
	\texttt{mvelezreyes@utep.edu} \\
}

\renewcommand{\shorttitle}{Smoothing Traffic Flow Through Automated Vehicle Control}

\begin{document}
\maketitle

\begin{abstract}
Stop-and-go traffic waves are known for reducing the efficiency of transportation systems by increasing traffic oscillations and energy consumption. In this study, we develop an approach to synthesize a class of additive feedback controllers for automated vehicles (AVs) to smooth nonlinear mixed traffic flow, including both AVs and human-driven vehicles (HVs). Unlike recent explicit AV controllers that rely on strict assumptions such as time-varying equilibrium traffic speed, our proposed AV controller requires only local traffic information, such as inter-vehicle spacing and relative speed, which are readily available through AV onboard sensors. Essentially, it allows a controlled AV to track a subtler version of the perturbed speed profile resulting from its preceding vehicle, thereby enabling smoother traffic flow. Additionally, we provide a method for selecting the optimal control parameters to achieve traffic-smoothing effects efficiently. These unique features of the developed AV controller ensure much higher implementability. We demonstrate the effectiveness of the proposed approach through simulations of two distinct traffic scenarios with varying levels of oscillation. The results show that AVs using the proposed controller are capable of effectively reducing traffic oscillations and lowering vehicle fuel consumption by up to 46.78\% and 2.74\%, respectively, for a platoon of 10 vehicles. The traffic-smoothing effect of the controller is more pronounced at higher penetration rates of AVs. While the performance of the proposed approach is slightly less superior to that of the most recent additive AV controller, it offers greater implementability and provides an efficient method for selecting optimal control parameters.
\end{abstract}


\section{Introduction}\label{section1}

Traffic congestion is a long-standing problem that has attracted significant research interest over the past few decades. While it can be triggered by perceptible factors such as lane changing~\citep{laval2006lane}, bottlenecks~\citep{kerner2012physics} and merging~\citep{milanes2010automated}, congestion often occurs even without these triggers. This is because small perturbations in traffic flow can propagate and amplify into noticeable stop-and-go waves due to traffic instabilities caused by the collective behavior of human drivers~\citep{flynn2009self}. Such unstable traffic has been observed in real-world experiments~\citep{sugiyama2008traffic}, leading to higher fuel consumption and emissions compared to smooth traffic flow~\citep{wu2019tracking}. 

To effectively regulate traffic flow and mitigate congestion, various traffic control methods have been developed, including signal control~\citep{ahmed2018optimum}, ramp metering~\citep{shang2023extending}, and variable speed limit (VSL) control~\citep{smulders1990control}. Signal control, one of the most widely used techniques, is primarily employed at intersections to regulate vehicular movements through the installation of traffic lights, a requirement shared with ramp metering. VSL control adjusts message signs based on real-time traffic information collected by sensors deployed along roadways. While these conventional techniques have proven effective in managing traffic flow, they rely heavily on fixed infrastructure, such as traffic signals and roadside sensors, which limits their flexibility.

In addition to the aforementioned conventional traffic control approaches, recent advancements in self-driving technologies have introduced new possibilities for future traffic operations and control. Automated vehicles (AVs) can now act as mobile actuators within the traffic flow, enabling Lagrangian traffic control~\citep[Chapter 5]{bayen2022control}. This method effectively smooths unstable and oscillatory mixed traffic, involving both AVs and human-driven vehicles (HVs), by controlling a small number of AVs to improve overall traffic efficiency and performance. For instance, a linear optimal feedback AV controller designed in~\citep{zheng2020smoothing} aims to smooth traffic based on linearized car-following dynamics, assuming AVs have access to the state of all HVs in the flow. Notably, the assumption of knowledge about HVs is rather limited, and using car-following models linearized at flow equilibrium points does not necessarily capture the highly nonlinear nature of mixed traffic~\citep{zhu2018analysis}. Our prior work~\citep{wang2022optimal} synthesizes a nonlinear optimal additive AV controller to smooth mixed traffic without being limited to linearized dynamics. However, it does not provide an analytical proof of AV safety. Our recent study~\citep{wang2023general} improves upon this by developing a novel virtual tracking technique for an AV to follow a subtler version of the oscillatory speed from its preceding vehicle, with proven guarantees on safety and traffic-smoothing performance.

Additionally, machine learning techniques have been employed to develop driving strategies for AVs to smooth traffic flow. For example, reinforcement learning-based traffic-smoothing cruise controllers~\citep{lichtle2022deploying,lichtle2023traffic} have shown effectiveness in improving traffic energy efficiency. A Long Short-Term Memory (LSTM) network-based framework is developed in~\citep{li2024customizable} for AV control, considering physics-informed constraints, with an objective customizable to reduce vehicle speed variation. Assuming vehicle-to-vehicle (V2V) communication, the distributed data-driven predictive control strategy proposed in~\citep{wang2023distributed} enables connected automated vehicles (CAVs) to cooperatively mitigate traffic waves, thus enhancing traffic efficiency. Despite their effectiveness, these learning-based approaches require extensive real-world data for training appropriate AV controllers. Moreover, the trained controllers generally lack an analytical form with mathematical guarantees and may not perform reliably across diverse traffic conditions not covered in the training dataset.

Complementary to existing work, the present study focuses on synthesizing a class of additive AV controllers that can effectively smooth mixed traffic flow without requiring any linearization of its dynamics. The designed controllers have an explicit analytical form, which is useful for mathematical analysis and implementation. Additionally, the controller synthesis only requires local traffic information, which can be readily obtained through AV onboard sensors, without the need for V2V communication. We address two key limitations in our recent work~\citep{wang2023general}: 1) the requirement for equilibrium traffic speed in AV controller synthesis, and 2) the lack of systematic control parameter selection. Specifically, the main contributions of this work are summarized as follows:
\begin{itemize}
    \item We synthesize a class of additive AV controllers that can effectively smooth nonlinear mixed traffic flow without linearizing its dynamics. The designed controllers have an explicit analytical form and do not require any knowledge of the equilibrium traffic speed, unlike those presented in~\citep{wang2023general}. This significantly improves the implementability of the proposed AV control strategy. 
    
    \item We provide an effective computational method for selecting the optimal control parameter values for the synthesized additive AV controllers. Unlike~\citep{wang2023general}, where the AV control parameters are chosen intuitively, the present study offers a more systematic approach for implementation.
    
    \item We conduct comprehensive simulations to holistically evaluate the performance of the new additive AV controllers across various AV penetration rates, considering their impacts on both mobility and energy consumption of the traffic flow. This provides useful insights into future traffic operations and control with AVs. 
\end{itemize} 

The remainder of this article is outlined as follows. In Section~\ref{section2}, we present a mathematical framework to describe mixed traffic flow based on car-following dynamics. Subsequently, in Section~\ref{section3}, we synthesize a class of additive AV controllers with an explicit form and develop an efficient method for selecting the optimal control parameters. In Section~\ref{section4}, we conduct a series of numerical simulations to illustrate the effectiveness of the proposed approach in traffic smoothing. Finally, in Section~\ref{section5}, we conclude the article and discuss future research directions.

\section{Mixed Traffic Flow Dynamics}\label{section2}

\begin{figure}[t!]
    \centering
    \includegraphics[width=0.8\linewidth]{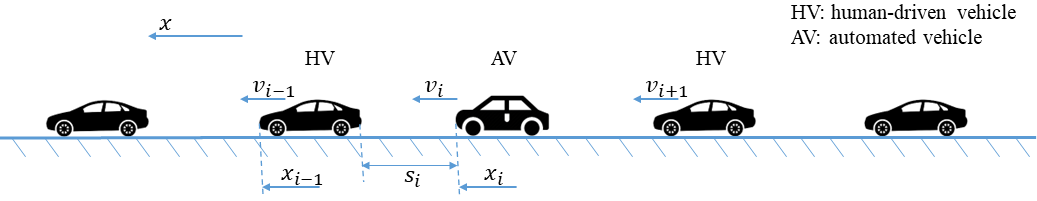}
    \vskip-6pt
    \caption{Illustration of mixed traffic flow involving automated vehicles (AVs) and human-driven vehicles (HVs). There is no specific requirement for the position of AVs within the traffic; this is for illustration purposes only.}
    \vskip-6pt
    \label{MixedTraffic}
\end{figure}

We consider a generic setting of mixed traffic flow consisting of AVs and HVs, as projected to be the case for the foreseeable future~\citep{wang2022policy}. As shown in Figure~\ref{MixedTraffic}, without loss of generality, a string of $N$ vehicles, denoted by the totally ordered set ${\cal N} = \{1, 2, \cdots, N\}$, is considered in the form of car-following dynamics. Let $x_{i}(t)$ and $v_{i}(t)$ denote the displacement and speed of any vehicle $i \in \cal{N}$ at time $t$, respectively. $s_{i}(t)$ signifies the inter-vehicle spacing between vehicle $i$ and its preceding vehicle $i-1$, defined as $s_{i}(t) = x_{i-1}(t) - x_{i}(t) - l_{i-1}$ with $l_{i-1}$ being the length of vehicle $i-1$. These notations are widely used in the literature to describe car-following dynamics, where one vehicle follows the driving behavior of its preceding vehicle, as detailed in~\citep{wilson2011car}. Lane-changing maneuvers are not considered to simplify the analysis, a common practice in studies focused on smoothing and stabilizing traffic flow through AV control~\citep{cui2017stabilizing,wu2018stabilizing,stern2018dissipation,zheng2020smoothing,giammarino2020traffic,wang2022optimal,wang2023general}. This assumption is reasonable because, in the context of traffic smoothing, imperfections such as lane changes and driver heterogeneity are considered external perturbations to a deterministic single-lane model, on which mathematical analysis is based~\citep{wilson2011car}. For brevity in the analysis, the time argument $t$ is omitted hereafter.

The motion of any vehicle, $i \in \cal{N}$, is described by the following ordinary differential equations~\citep{wilson2011car}
\begin{eqnarray}
    &~&\dot{x}_{i} = v_{i},    \label{eq1.1}   \\
    &~&\dot{v}_{i} = f(s_{i}, \Delta v_{i}, v_{i}),     \label{eq1.2}
\end{eqnarray}
where the nonlinear function $f$ describes the relationship between the acceleration of the $i$th vehicle, $\dot{v}_{i}$, and the car-following variables $s_{i}$, $\Delta v_{i}$, and $v_{i}$. $\Delta v_{i}$ denotes the relative speed between two consecutive vehicles, defined as
\begin{eqnarray}
    \Delta v_{i} = \dot{s}_{i} = v_{i-1} - v_{i}.     \label{eq1.3}
\end{eqnarray}
Equations~\eqref{eq1.1}--\eqref{eq1.2} are widely adopted in the literature to represent the general functional form of car-following dynamics, where speed and acceleration are assumed to be physically bounded. 

Following Figure~\ref{MixedTraffic}, let $\cal{H}$ and $\cal{A}$ denote the totally ordered set of HVs and AVs, respectively, such that $\cal{H} \cup \cal{A} = \cal{N}$. HVs are assumed to follow generic car-following dynamics given by equations~\eqref{eq1.1}--\eqref{eq1.2}. For AVs, our objective is to synthesize a general class of additive feedback controllers to smooth mixed traffic flow in the presence of oscillations. To this end, the dynamics of mixed traffic are described as follows
\begin{eqnarray}
    \dot{x}_{i} = v_{i}, ~\forall~ i \in \cal{N}    \label{eq1.4}
\end{eqnarray}
\begin{numcases}{\dot{v}_{i} =}
    g(s_{i}, \Delta v_{i}, v_{i}), ~\forall~ i \in \cal{H}   \label{eq1.5a}  \\  [3pt]
    h(s_{i}, \Delta v_{i}, v_{i}) + u_{i}, ~\forall~ i \in \cal{A}    \label{eq1.5b}
\end{numcases}
where $u_{i}$ represents the additive control input that needs to be characterized, while the actual AV acceleration is determined by equation~\eqref{eq1.5b} in its entirety. According to equation~\eqref{eq1.5a}, controlled AVs do not alter the operation of HVs, as they continue to adhere to regular car-following principles. It is noted that the functionals $g$ and $h$ do not necessarily need to be identical; instead, they depend on the specific car-following principles followed by HVs and AVs. This form of additive control input for AVs, shown in equation~\eqref{eq1.5b}, has been adopted in various studies~\citep{wang2022optimal,wang2022smoothing,delle2022new,wang2023general}, ensuring compliance with car-following principles for AVs.

In equilibrium traffic, all vehicles drive at a uniform speed~\citep{cui2017stabilizing}, satisfying $\dot{v}_{i} = 0$ in equations~\eqref{eq1.5a} and~\eqref{eq1.5b}. In this state, no additional control input $u_{i}$ is necessary for AVs, as the equilibrium is maintained solely by adhering to car-following principles. However, when stop-and-go waves emerge and approach a controlled AV, $u_{i}$ will act to dampen the undesired traffic waves, reducing the disturbance experienced by the vehicles behind. One effective way to achieve this is by controlling an AV to closely track a subtler version of the perturbation propagating from its preceding vehicle, as first introduced in~\citep{wang2023general}. In the subsequent section, we will derive an explicit form of a class of additive controllers, $u_{i}$, for AVs to achieve these goals, thereby alleviating traffic oscillations and smoothing the flow.

\section{Synthesizing Explicit AV Controllers for Traffic Smoothing}\label{section3}

In this section, we synthesize a class of additive AV controllers in a general functional form for traffic smoothing. These controllers enable an AV to track a specifically designed speed profile, leading to reduced perturbations caused by traffic waves. Furthermore, we derive an explicit form of the synthesized additive controllers for illustration and present a method for determining the optimal control parameter values.

\subsection{Synthesizing a Class of Additive AV Controllers}\label{section3.1}

Considering the mechanism of traffic wave propagation, our goal is to design a class of additive AV controllers that enable a controlled AV to closely follow a virtual speed profile for smoothing traffic flow. This design ensures that speed perturbations are effectively reduced when passing a controlled AV compared to the case if it were a HV. As a result, backward-propagating traffic waves are mitigated, leading to smoother traffic flow. Notably, we synthesize the AV controller using readily available local traffic information only, including inter-vehicle spacing $s_{i}$ and relative speed $\Delta v_{i}$, which enhances implementability. Moreover, the virtual speed profile designed for an AV to follow essentially represents a subtler version of the disturbance resulting from its immediate preceding vehicle. To ensure car-following safety for the controlled AV, we derive sufficient conditions to guide the selection of optimal control parameters. Specifically, we synthesize a class of additive controllers as follows
\begin{eqnarray}
    u_{i} = \mathcal{I}(s_{i}, \Delta v_{i}),    \label{eq3.1}
\end{eqnarray}
where the function $\mathcal{I}: \mathbb{R}_{+} \times \mathbb{R} \longrightarrow \mathbb{R}$ satisfies the following conditions:

(i) $\mathcal{I}(s_{i}, \Delta v_{i})$ is strictly monotonically increasing in $s_{i}$ and $\Delta v_{i}$;

(ii) $\mathcal{I}(s_{i}, \Delta v_{i}) \cdot \Delta v_{i} > 0$ when $\Delta v_{i} \neq 0$, and $\mathcal{I}(s_{i}, \Delta v_{i}) = 0$ when $\Delta v_{i} = 0$;

(iii) $\mathcal{I}(s_{i}, \Delta v_{i})$ is twice differentiable in $t$ with bounded derivatives;

(iv) $\mathcal{I}(s_{i}, \Delta v_{i})$ is bounded from above by a positive real number, say $\alpha > 0$, i.e., $\sup_{s_{i},\Delta v_{i}} \mathcal{I}(s_{i}, \Delta v_{i}) = \alpha$.

As mentioned before, $s_{i}$ and $\Delta v_{i}$ are functions of time; hence, $\mathcal{I}$ is an implicit functional of $t$, which is not explicitly written out for ease of notation. Following a similar principle shown in~\cite[Theorem 3.1]{wang2023general}, one can verify that the speed of the controlled AV, $v_{i}$, converges to $\tilde{v}_{i} = v_{i-1} + \mathcal{I}(s_{i}, \Delta v_{i})$, termed the virtual speed profile, in finite time. Essentially, $\tilde{v}_{i}$ provides the AV with a subtler version of the perturbed speed of its preceding vehicle to follow when the above conditions are satisfied by the function $\mathcal{I}$. Consequently, undesired traffic waves can be mitigated to some extent as they propagate through the controlled AV, resulting in smoother traffic flow. 

We present below some brief discussions on the conditions~(i)--(iv) listed above.

(a) Condition~(i) suggests that a controlled AV is likely to accelerate more as the spacing ($s_i$) and relative speed ($\Delta v_i$) to the vehicle ahead increase, aligning with the rational driving constraints (RDC)~\citep{wilson2011car} of car-following dynamics.

(b) According to the RDC~\citep{wilson2011car}, condition~(ii) implies that a controlled AV tends to accelerate ($\mathcal{I} > 0$) when its speed is lower than that of the vehicle in front ($\Delta v_{i} > 0$) in car-following scenarios. Conversely, it is likely to decelerate when traveling faster than its preceding vehicle. Moreover, the additive control $u_i$ remains inactive ($\mathcal{I} = 0$) when there is no relative speed difference ($\Delta v_i = 0$) between the two vehicles.

(c) Condition~(iii) aids in ensuring the convergence of $v_{i}$ to $\tilde{v}_{i}$, for virtual speed tracking. The proof is omitted here as it is similar to that of~\cite[Theorem 3.1]{wang2023general}.

(d) Condition~(iv) is established to ensure safety for the controlled AV by preventing it from following an excessively aggressive virtual speed profile. Essentially, the AV's tracked speed profile should be closely aligned with that of the vehicle immediately ahead. In the following subsection, we will provide a brief analysis to derive sufficient conditions for $\alpha$, using an explicit example of the function $\mathcal{I}$. 

\begin{figure}[t!]
    \centering
    \includegraphics[width=0.6\linewidth]{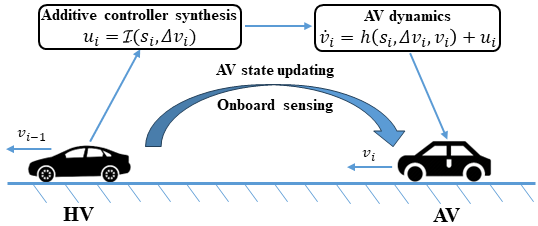}
    \vskip-6pt
    \caption{Graphic representation of the AV controller synthesis. The proposed method is independent of the type of vehicle immediately ahead; an HV is depicted as the leader for demonstration purposes only. The controlled AV acquires the necessary traffic information, including $s_i$ and $\Delta v_i$, through its onboard sensors. Using this information, the additive controller $u_i$ is synthesized based on equation~\eqref{eq3.1} to calculate the acceleration control input according to equation~\eqref{eq1.5b}, allowing the AV to update its dynamics.}
    \vskip-6pt
    \label{controller_synthesis}
\end{figure}  

Figure~\ref{controller_synthesis} presents a schematic illustrating the mechanism of AV controller synthesis. The controlled AV acquires traffic information, including spacing ($s_i$) and relative speed ($\Delta v_i$), through its onboard sensors. Using this information, the additive AV controller $u_{i} = \mathcal{I}(s_{i}, \Delta v_{i})$ is synthesized to enable the AV to closely track the virtual speed profile $\tilde{v}_{i} = v_{i-1} + \mathcal{I}(s_{i}, \Delta v_{i})$. Once $u_i$ is synthesized, acceleration control is executed according to equation~\eqref{eq1.5b} to update the AV's dynamics. This process is continuous, with the controlled AV consistently acquiring local traffic information relative to its preceding vehicle as it moves forward.

\begin{remark}\label{Remark1}
In~\citep{wang2023general}, we developed a first-of-its-kind additive AV controller aimed at synthesizing effective AV controls for smoothing mixed traffic flow without linearizing nonlinear car-following dynamics. A notable limitation of the AV controller proposed in~\citep{wang2023general} is its dependence on the knowledge of the equilibrium traffic speed. This equilibrium speed can be challenging to determine, as it varies with traffic density over time~\citep{shang2023extending}, thus limiting the practicality of the approach. In contrast, the class of additive controllers developed in this study is more general, eliminating the need for the equilibrium speed requirement. The necessary information, $s_{i}$ and $\Delta v_{i}$, is readily available through an AV's onboard sensors, enhancing the method's implementability. 
\end{remark}

\begin{remark}\label{Remark2}
The function $\mathcal{I}$ described above represents a broad class of additive AV controllers. It can be verified that a series of sigmoid functions satisfy the conditions~(i)--(iv) and can be utilized to synthesize the controller $u_{i}$. Some useful examples include the inverse tangent function $\arctan(\cdot)$, the hyperbolic tangent function $\tanh(\cdot)$, and the Gauss error function $erf(\cdot)$.
\end{remark}

\subsection{An Explicit Form of the Additive AV Controller}\label{section3.2}

While a general class of additive controllers for AVs has been derived in Section~\ref{section3.1}, an explicit form of the function $\mathcal{I}$ is necessary for simulation and implementation. Here we present such an example for illustration. Specifically, 
\begin{eqnarray}
    \mathcal{I}(s_{i}, \Delta v_{i}) = \beta \cdot \arctan(\gamma s_{i} \Delta v_{i}),  \label{eq3.2}
\end{eqnarray}
where $\beta$ and $\gamma$ are non-negative control parameters. It is easy to verify that the function $\mathcal{I}$ given by equation~\eqref{eq3.2} satisfies the conditions~(i)--(iv) presented in Section~\ref{section3.1}. Considering equation~\eqref{eq3.2}, it is evident that its upper limit is $\alpha = (\beta\pi)/2$.  

For any given $\mathcal{I}$, it is important to ensure car-following safety. This depends on $\alpha$, which dictates the degree of subtlety of $\tilde{v}_{i}$ relative to the perturbed AV speed in the absence of the additive control input $u_i$. It is necessary that $\tilde{s}_{i}(t) \geq \check{s}$ for all $t \in I \coloneqq [0, t_{\textnormal{f}}]$ to ensure AV safety, where $t_{\textnormal{f}}$ is the terminal time. Here, $\tilde{s}_{i}(t)$ denotes the spacing between vehicle $i-1$ and the controlled AV $i$ at time $t$, with the AV following the virtual speed profile $\tilde{v}_{i}$, and $\check{s}$ represents the minimum safe spacing. It follows that 
\begin{align}
    \tilde{s}_{i}(t)
    & = s_{i}(0) + \int_{0}^{t}\left(v_{i-1}(\mu) - \tilde{v}_{i}(\mu)\right)d\mu   \nonumber  \\
    & = s_{i}(0) - \int_{0}^{t}\mathcal{I}(s_{i}(\mu),\Delta v_{i}(\mu))d\mu  \nonumber \\
    & > s_{i}(0) - \alpha t \geq s_{i}(0) - \alpha t_{\textnormal{f}},
    \label{eq3.3}
\end{align}
where $s_{i}(0)$ is the initial safe spacing between the controlled AV $i$ and its preceding vehicle $i-1$. It follows from~\eqref{eq3.3} that
\begin{eqnarray}
    \tilde{s}_{i}(t) > s_{i}(0) - \alpha t_{\textnormal{f}} \geq \check{s},
    \label{eq3.4}
\end{eqnarray}
which leads to
\begin{eqnarray}
    \alpha = \frac{\beta\pi}{2} \leq \frac{s_{i}(0)-\check{s}}{t_{\textnormal{f}}} \Longrightarrow \beta \leq \frac{2[s_{i}(0)-\check{s}]}{\pi t_{\textnormal{f}}}.  \label{eq3.5}
\end{eqnarray}

\begin{remark}\label{Remark3}
The upper bound on $\beta$ derived above adheres to car-following principles. A larger initial spacing $s_{i}(0)$ provides the AV with more flexibility to adjust its speed, thereby leading to a greater upper bound on $\beta$. Conversely, increased traffic perturbations over a longer driving horizon $t_{\textnormal{f}}$ may limit the available space to construct a virtual speed profile for the AV to track, resulting in a smaller upper bound on $\beta$.
\end{remark}

\begin{remark}\label{Remark4}
The upper bound on $\beta$ given by~\eqref{eq3.5} is derived conservatively. This approach considers the worst-case scenario where traffic perturbations persist throughout the entire time horizon $I$. Consequently, the derived conditions ensure AV safety for any duration of speed perturbations within $I$. Clearly, a larger upper bound on $\beta$ can be derived for scenarios with a speed perturbation period shorter than $t_{\textnormal{f}}$.
\end{remark}

\subsection{Optimal Control Parameter Selection}\label{section3.3}

While the results in Section~\ref{section3.2} provide a range of options for $\beta$, our focus is on selecting optimal values for the non-negative control parameters $\beta$ and $\gamma$, adhering to the derived conditions. As shown in~\citep{wang2022smoothing}, reducing the speed difference between an AV and its preceding vehicle effectively mitigates traffic oscillations, thereby smoothing the flow. Following this principle, we formulate an optimal control problem to determine these values, aiming for the controlled AV to closely mimic the speed of its preceding vehicle for traffic smoothing. Hence, we define the following objective function to minimize
\begin{eqnarray}
    J = \frac{1}{2} \int_{0}^{t_{\textnormal{f}}} (v_{i} - v_{i-1})^2dt.  \label{eq3.6}
\end{eqnarray}
The goal is to minimize this function to determine the optimal values for the non-negative control parameters, $\beta$ and $\gamma$, subject to~\eqref{eq3.5}. 

This problem is analogous to the least squares estimation of system parameters described in~\citep{ahmed1976simple}. Due to the time-invariant nature of the control parameters, the standard maximum principle approach is not applicable. Following the gradient technique proposed in~\citep{ahmed1976simple}, we outline an iterative computational algorithm for determining the optimal values of $\beta$ and $\gamma$. 

For any controlled AV, let us rewrite~\eqref{eq1.5b} as follows  
\begin{eqnarray}
\dot{v}_{i} = h(s_{i}, \Delta v_{i}, v_{i}) + u_{i} \coloneqq r(s_{i}, \Delta v_{i}, v_{i}, \theta),  \label{eq3.7}
\end{eqnarray}
where $u_i = \beta \cdot \arctan(\gamma s_{i} \Delta v_{i})$, and $\theta = [\beta, \gamma]^\prime$ with the prime denoting the transpose. 

Since we are employing a gradient-based iterative computational method, it is necessary to establish the descent direction for each iteration. Following the approach outlined in~\citep{ahmed1976simple}, we define a vector $z = [z_1, z_2]^\prime$, whose dynamics are governed by 
\begin{eqnarray}
\dot{z} = \frac{\partial r}{\partial v_i} z + \frac{\partial r}{\partial \theta}.  \label{eq3.8}
\end{eqnarray}
By employing this basic notation change, one can solve equation~\eqref{eq3.8} to find $z(t)$ and compute the descent direction as follows
\begin{eqnarray}
    \lambda_{\theta} = \int_{0}^{t_\textnormal{f}}z(t)[v_{i}(t) - v_{i-1}(t)]dt, \label{eq3.9}
\end{eqnarray}
for $\beta$ and $\gamma$ at each iteration. In the subsequent section, we will present a detailed step-by-step computational procedure for determining the optimal values of these control parameters.

\section{Numerical Results}\label{section4}

In this section, we demonstrate the effectiveness of the proposed additive AV controller in smoothing mixed traffic flow through numerical simulations. First, we introduce explicit car-following models for both HVs and AVs, the abstract forms of which were presented in Section~\ref{section2}. Subsequently, we outline a step-by-step computational procedure for calculating optimal values of $\beta$ and $\gamma$. Finally, we present extensive simulation results to illustrate the efficacy of the proposed AV controls. 

\subsection{Illustration of Car-Following Dynamics}\label{section4.1}

Following the modeling framework for mixed traffic flow presented in Section~\ref{section2}, we introduce concrete car-following models here to characterize the dynamics of HVs and AVs for numerical studies. Specifically, we employ the widely used intelligent driver model (IDM)~\citep{treiber2000congested} for HVs, which has been extensively utilized in the literature~\citep{treiber2013traffic,wu2018stabilizing,wang2022optimal,wang2022smoothing,shang2023extending,wang2023general} due to its accurate representation of human driving behavior. 

Subsequently, the HV dynamics given by equation~\eqref{eq1.5a} are explicitly written as
\begin{eqnarray}
    g = a \left[ 1 - \left(\frac{v_{i}}{v_{0}}\right)^{\delta} - \left(\frac{s^{\ast}(v_{i},\Delta v_{i})}{s_{i}}\right)^{2} \right], ~ i \in \mathcal{H},   \label{eq4.1}
\end{eqnarray}
where
\begin{equation}
    s^{\ast}(v_{i},\Delta v_{i}) = s_{0} + \max\left\{0, v_{i}T - \frac{v_{i}\Delta v_{i}}{2\sqrt{ab}}\right\},    \label{eq4.2}
\end{equation}
with the model parameter values presented in Tables~\ref{Table_parameters_ScenarioI} and~\ref{Table_parameters_ScenarioII} for different scenarios of the numerical study.

As seen in the literature~\citep{gunter2019modeling,wang2022optimal}, the optimal velocity with relative velocity (OVRV) model~\citep{milanes2013cooperative} is widely employed to describe the car-following behavior of AVs. This model has found extensive application in (cooperative) adaptive cruise control (ACC) systems, the first generation of AVs, and has shown high accuracy in simulating both ACC trajectories and real-world scenarios~\citep{liang1999optimal,milanes2013cooperative,shang2024two}. Following the OVRV model, the function $h$ in equation~\eqref{eq1.5b} is expressed as
\begin{eqnarray}
    h = k_{1}\left(s_{i} - \eta - \tau v_{i}\right) + k_{2}\Delta v_{i}, ~ i \in \mathcal{A},       \label{eq4.3}
\end{eqnarray}
with the model parameters shown in Table~\ref{Table_parameters_ScenarioI}. Therefore, the AV dynamics with an additive control input are written as
\begin{eqnarray}
    h = k_{1}\left(s_{i} - \eta - \tau v_{i}\right) + k_{2}\Delta v_{i} + u_i, ~ i \in \mathcal{A},       \label{eq4.4}
\end{eqnarray}
with $u_i$ given by equation~\eqref{eq3.2}. 

\subsection{Computational Procedure for Optimal Parameter Selection}\label{section4.2}

While Section~\ref{section3.3} briefly introduced a method for selecting optimal values of the control parameters $\beta$ and $\gamma$, following the approach developed in~\citep{ahmed1976simple}, here we present a detailed step-by-step computational procedure for numerical implementation. 

\vskip3pt 
Step 1: Initialize the state of the car-following system and begin the first iteration $\kappa=1$ with an initial feasible $\theta$, continuing until the last iteration $N_{\text{max}}=300$. Let $\theta^{(\kappa)}$ denote the value of $\theta$ at the $\kappa$th iteration, starting from $\theta^{(1)}$.

Step 2: Solve equation~\eqref{eq3.7} to obtain the pair $(v_i^{(\kappa)}, \theta^{(\kappa)})$, and calculate the value of $J^{(\kappa+1)}$ using equation~\eqref{eq3.6}.

Step 3: Solve equation~\eqref{eq3.8} with $z(0) = 0$ to find $z^{(\kappa)}$. 

Step 4: Compute equation~\eqref{eq3.9} to obtain the descent direction, $\lambda_{\theta}^{(\kappa)}$.

Step 5: If $\lambda_{\theta}^{(\kappa)} = 0$, then $\theta^{(\kappa)}$ corresponds to the position of the minimum for the functional $J$. Otherwise, the control parameter vector is updated as $\theta^{(\kappa+1)} = \theta^{(\kappa)} - \epsilon \lambda_{\theta}^{(\kappa)}$, where $\epsilon \in (0,1)$ is a small positive step size, ensuring that $\theta^{(\kappa+1)}$ remains feasible. 

Step 6: Compute the value of $J^{(\kappa+1)}$ and check if 
\begin{eqnarray}
\left|J^{(\kappa+1)} - J^{(\kappa)}\right| > \phi   \label{eq4.5}
\end{eqnarray}
holds for a prespecified small positive number $\phi$. If so, proceed to Step 2 with $\theta^{(\kappa)}$ replaced by $\theta^{(\kappa+1)}$ and repeat the procedure. The process continues until $\left|J^{(\kappa+1)} - J^{(\kappa)}\right| < \phi$, or until the maximum number of iterations, $N_{\text{max}}$, is reached. If either of the condition is not met, the computational procedure terminates. 

\subsection{Simulation Results}\label{section4.3}

We conduct extensive numerical simulations in~\texttt{MATLAB} to demonstrate the effectiveness of the proposed approach in smoothing traffic flow and enhancing overall traffic performance. As previously introduced, the IDM and OVRV model are utilized to simulate the dynamics of HVs and AVs, respectively, with AVs benefiting from the additive controller using optimal control parameter values. It is noted that our proposed control method effectively operates for each individual AV without requiring inter-vehicle communication. Similar to~\citep{wang2022optimal,wang2023general,aguilar2024energy}, we evaluate a platoon consisting of 10 vehicles following a lead vehicle executing a specified speed profile, facilitating variations in the market penetration rate (MPR) of AVs. AVs are evenly distributed within the platoon, without considering parameter heterogeneity, as seen in~\citep{jin2014dynamics,wang2023general,aguilar2024energy}. This assumption is reasonable since the impact of our proposed AV control strategy on traffic flow is evaluated using average metrics.

\begin{table}[t!]
\setlength{\tabcolsep}{0.8pt}
\caption{Model parameter values of the IDM~\citep{treiber2013traffic,de2021calibrating} and OVRV model~\citep{gunter2019modeling,wang2023general}}\label{Table_parameters_ScenarioI}
\vspace*{-2mm}
\begin{center}
 \begin{tabular}{c c c c c c c c}
 \hline
 \textbf{IDM} ~&~ $a$ (m/s\textsuperscript{2}) ~&~ $b$ (m/s\textsuperscript{2}) ~&~ $v_{0}$ (m/s) ~&~ $s_{0}$ (m) ~&~ $T$ (s) ~&~ $\delta$ ~&~ $l_{i}$ (m)  \\  [0.5ex]
 \textbf{Value} ~&~ 0.6 ~&~ 2.5 ~&~ 35 ~&~ 2 ~&~ 1.5 ~&~ 4 ~&~ 5 \\ [0.3ex]
 \hline
 \hline
 \textbf{OVRV} ~&~ $k_{1}$ (s\textsuperscript{-2}) ~&~ $k_{2}$ (s\textsuperscript{-1}) ~&~ $\eta$ (m) ~&~ $\tau$ (s) ~&~ $l_{i}$ (m) ~&~ ~&~  \\  [0.5ex]
 \textbf{Value} ~&~ 0.02 ~&~ 0.13 ~&~ 21.51 ~&~ 1.71 ~&~ 5 ~&~ ~&~    \\ [0.3ex]
 \hline
\end{tabular}
\end{center}
\end{table}

\begin{figure}[t!]
 	\centering
 	\subfloat[\textnormal{MPR = 0\%}]{\includegraphics[width=0.32\textwidth]{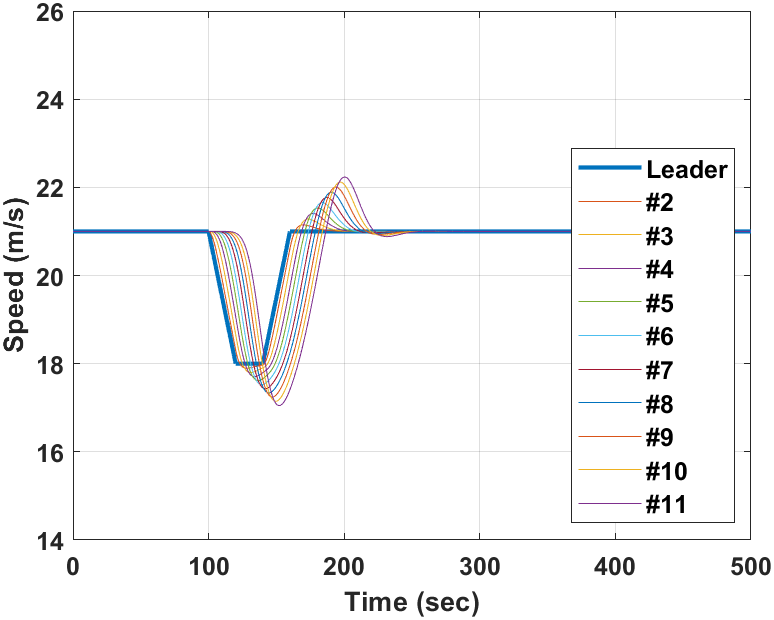}\label{Speed_0_Scenario1}}
 	\hfil%
 	\subfloat[\textnormal{MPR = 50\%}]{\includegraphics[width=0.32\textwidth]{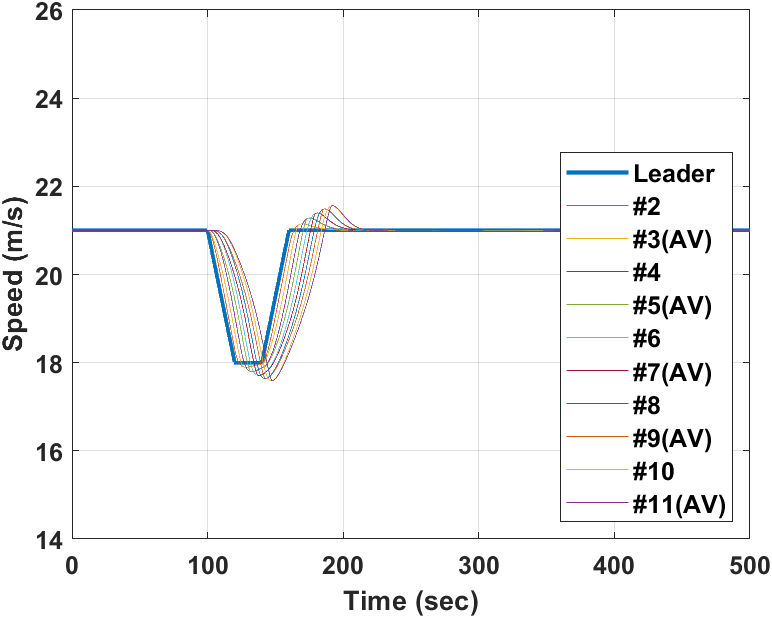}\label{Speed_50_Scenario1}}
 	\hfil%
 	\subfloat[\textnormal{MPR = 100\%}]{\includegraphics[width=0.32\textwidth]{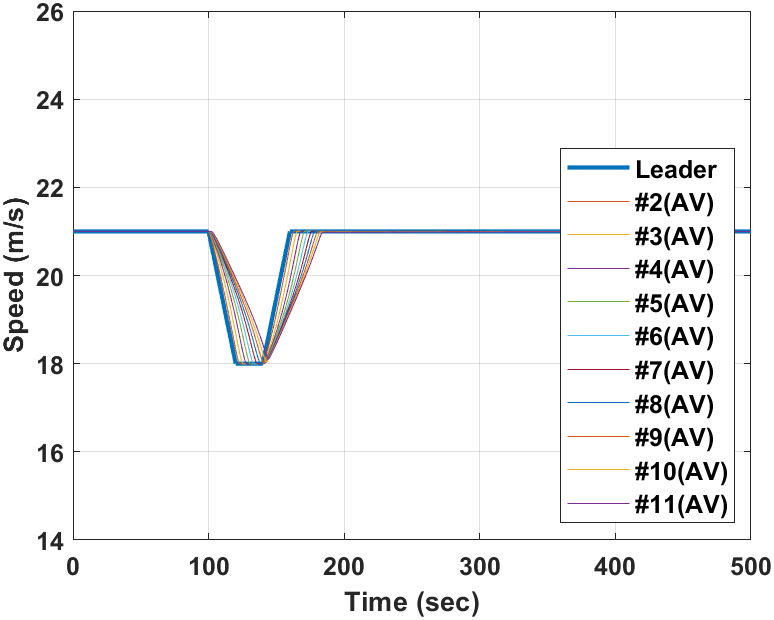}\label{Speed_100_Scenario1}}
    \caption{Scenario~I: 2D plot of vehicle speeds under MPR = 0\%, 50\%, and 100\%. AVs are controlled with the TS-OPS developed in this study, while HVs follow the IDM with parameter values shown in Table~\ref{Table_parameters_ScenarioI}. As the MPR of AVs increases, stop-and-go traffic waves decrease, leading to smoother traffic flow.}\label{Speed_ScenarioI}
\end{figure}

\begin{figure}[t!]
 	\centering
     \subfloat[\textnormal{MPR = 0\%}]{\includegraphics[width=0.33\textwidth]{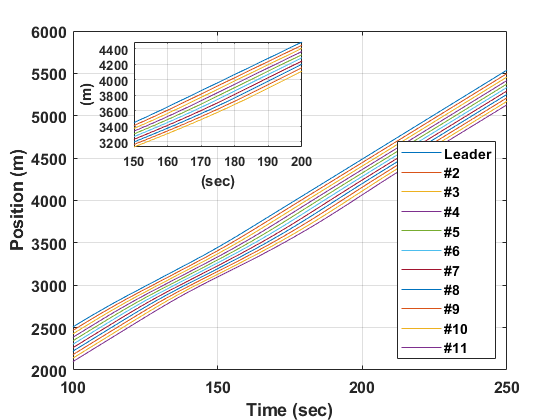}\label{Position_0_Scenario1}}
 	\hfil%
 	\subfloat[\textnormal{MPR = 50\%}]{\includegraphics[width=0.33\textwidth]{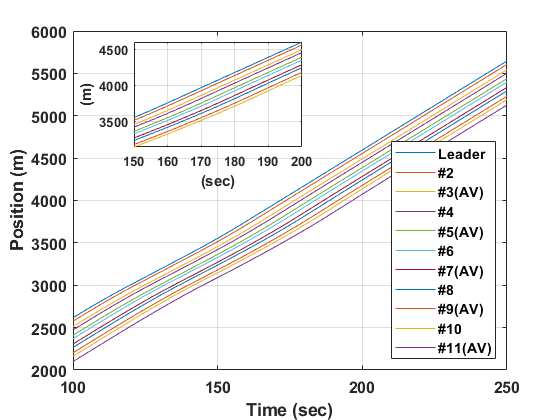}\label{Position_50_Scenario1}}
 	\hfil%
 	\subfloat[\textnormal{MPR = 100\%}]{\includegraphics[width=0.33\textwidth]{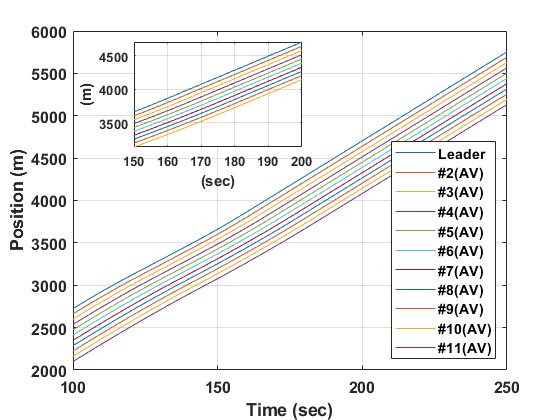}\label{Position_100_Scenario1}}
    \caption{Scenario~I: 2D plot of vehicle trajectories (positions) under MPR = 0\%, 50\%, and 100\%, corresponding to the speed profiles presented in Figure~\ref{Speed_ScenarioI}. Traffic becomes smoother as more AVs are introduced into the platoon, and car-following safety is ensured with no rear-end collisions observed.} %
    \label{Safety_ScenarioI}
\end{figure}

In addition to examining vehicle speed and trajectory profiles, we also utilize average speed variation (ASV) and fuel consumption (FC) metrics to measure the performance of the AV control strategy. ASV, defined as $\text{ASV}_i = \frac{1}{t_2-t_1} \int_{t_1}^{t_2}|v_i(t) - v^*|dt$ for vehicle $i$ over the period $[t_1, t_2]$, with $v^*$ being the desired (equilibrium) speed~\citep{wang2023general}, quantifies the deviation of vehicle speeds from a desired value, providing a measure of traffic smoothing. A higher ASV typically indicates greater traffic oscillations and reduced efficiency. FC is assessed using the VT-Micro model~\citep{ahn2002estimating}, a well-established method for estimating fuel consumption based on a vehicle's instantaneous speed and acceleration profiles using regression models. Interested readers can find detailed information about ASV and the VT-Micro model in~\citep{wang2023general} and~\citep{ahn2002estimating}, respectively. Both metrics are evaluated within a specified time window for the 10 vehicles behind the leader to eliminate the constant results from the leader and better capture the behavior of the following vehicles. These metrics collectively provide useful insights into the effectiveness of the proposed AV controller in improving traffic efficiency.

In what follows, two distinct scenarios are simulated to examine and quantify the impact of the proposed AV control strategy. Events in Scenarios~I and~II are simulated for a total of 500 seconds. The platoon starts in equilibrium traffic, with the leader driving at a constant speed of 21~m/s for 100 seconds. After this time has passed, the leader brakes for 20 seconds until it reaches a speed of 18~m/s, which it maintains for the next 20 seconds. The leader then accelerates for 20 seconds to resume the initial speed of 21~m/s, which it maintains until the end of the simulation. This speed profile, as shown in Figure~\ref{Speed_0_Scenario1}, is purposefully synthesized to generate oscillations and introduce instability into the vehicle trajectories~\citep{wang2022optimal,wang2023general,aguilar2024energy}. While the leader's trajectory remains the same for both scenarios, the simulations differ in the set of parameter values used for the IDM describing human driving behavior. In addition to holistically evaluating the effectiveness of the proposed traffic-smoothing controller with optimal parameter selection (termed TS-OPS hereafter), its performance is also compared against the most recent additive AV controller developed in~\citep{wang2023general} (termed TS-TRC in this study). The TS-TRC is given by $u_i = \varphi_{1}(\Delta v_i + \varphi_{2} \arctan[\varphi_{3} s_i (v^* - v_{i-1})])$, where $\varphi_{1}$, $\varphi_{2}$, and $\varphi_{3}$ are control parameters, and $v^*$ is the equilibrium traffic speed. As discussed in Section~\ref{section3}, the synthesis for TS-TRC suffers from a major limitation of requiring the knowledge of $v^*$, which typically varies with traffic density over time and can be challenging for vehicles to obtain, rendering it impractical for real-world implementation. Furthermore,~\citep{wang2023general} did not provide an effective method for selecting optimal values for the control parameters $\varphi_{1}$, $\varphi_{2}$, and $\varphi_{3}$.

\subsubsection{Scenario~I}\label{section4.3.1}

In this scenario, the parameter values adopted for the IDM and OVRV model are summarized in Table~\ref{Table_parameters_ScenarioI}, similar to those used in many prior studies~\citep{treiber2013traffic,de2021calibrating,gunter2019modeling,wang2023general}. The set of parameters for the IDM is used to simulate the driving behavior of HVs, exhibiting low levels of oscillation. Metrics to quantify AV impacts and traffic performance are calculated over the time interval of 100 to 250 seconds to eliminate constant data during other periods of the simulation. Figure~\ref{Speed_0_Scenario1} shows the speed profiles of all vehicles in the absence of AVs, i.e., MPR = 0\%. It is observed that speed perturbations amplify along upstream of the traffic, growing into stop-and-go waves. Once AVs with the TS-OPS controller are introduced into the traffic mixture, as demonstrated in Figures~\ref{Speed_50_Scenario1} and~\ref{Speed_100_Scenario1}, vehicle speed oscillations are reduced significantly with no overshoot or undershoot observed at 100\% MPR, demonstrating the capability of the proposed controller in smoothing traffic. The corresponding vehicle trajectory profiles are illustrated in Figure~\ref{Safety_ScenarioI}, showing no rear-end collisions. This demonstrates car-following safety, consistent with the results derived in Section~\ref{section3.2}.

 \begin{figure}[t!]
    \centering
    \includegraphics[width=0.4\linewidth]{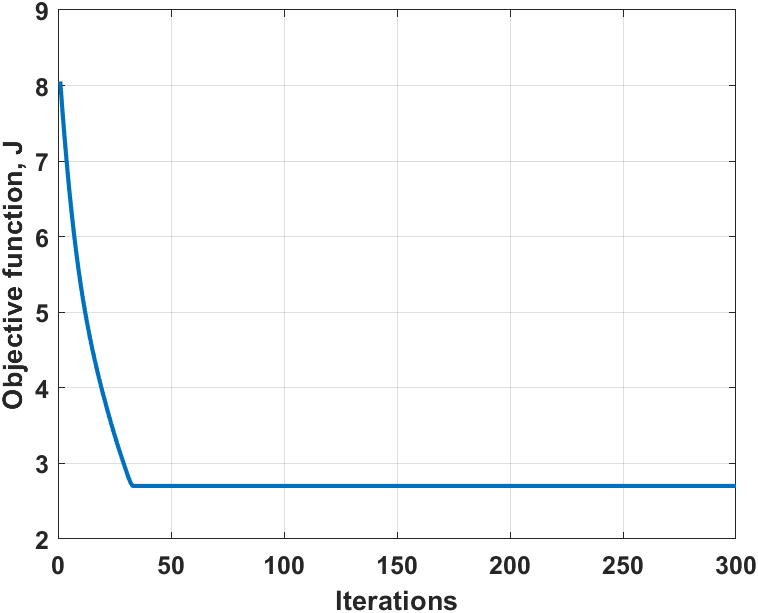}
    \vskip-6pt
    \caption{Scenario~I: The objective function $J$ converges as the number of iterations increases under a 10\% MPR.}
    \label{Objective_ScenarioI}
\end{figure}

\begin{figure}[t!]
    \centering
    \subfloat[\textnormal{ASV results at 10\% MPR}]{\includegraphics[width=0.4\linewidth]{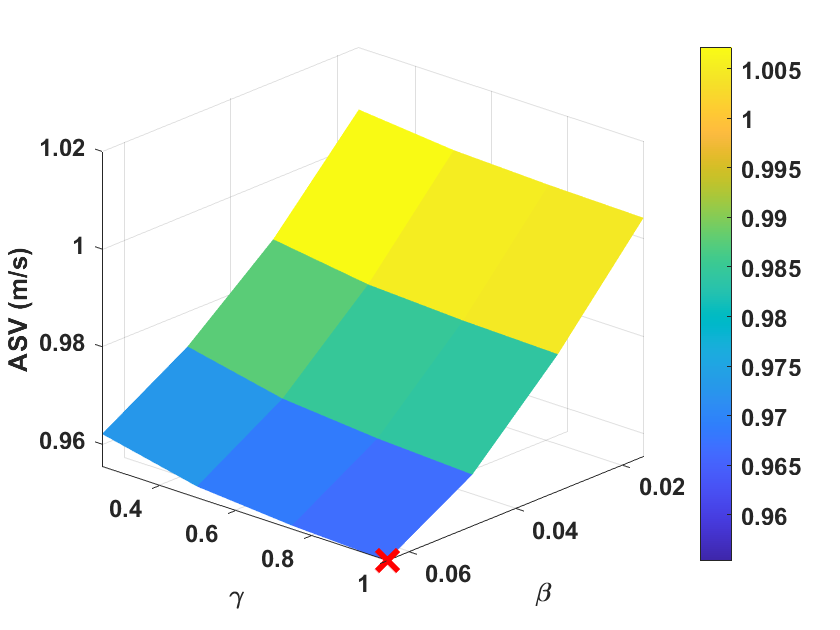}\label{ASV_Unoptimal_ScenarioI_10MPR}}
    \hfil%
    \subfloat[\textnormal{FC results at 10\% MPR}]{\includegraphics[width=0.4\linewidth]{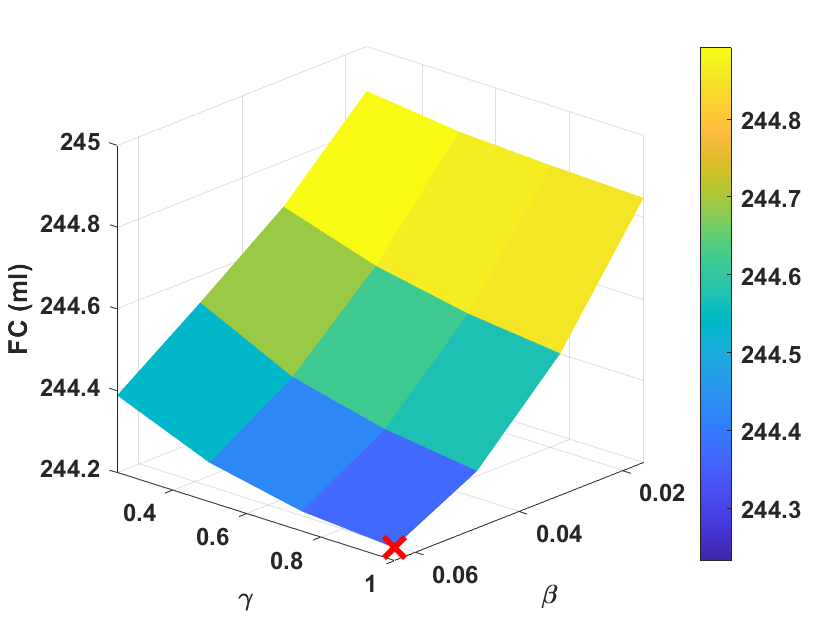}\label{FC_Unoptimal_ScenarioI_10MPR}}
    \caption{Scenario I: ASV and average FC results at 10\% MPR under multiple combinations of non-optimal $\beta$ and $\gamma$, considering $N_{\text{max}}=300$. The red marker denotes the optimal values of ASV and FC in Figures~\ref{ASV_Unoptimal_ScenarioI_10MPR} and~\ref{FC_Unoptimal_ScenarioI_10MPR}, respectively, achieved at the optimal control parameters, $\beta = 0.0642$ and $\gamma = 1.0011$. The optimal value of $\beta$ reaches its upper bound as determined by expression~\eqref{eq3.5}.}
    \label{ASV_FC_Unoptimal_ScenarioI_10MPR}
\end{figure}

As introduced in Section~\ref{section3.3}, one of the main characteristics of the proposed approach is its capability to determine the optimal set of control parameters, $\beta$ and $\gamma$, for each AV. This allows for an effective reduction in the speed discrepancy between the AV and its preceding vehicle. The algorithm presented in Section~\ref{section4.2} computes a new value of $\theta = [\beta, \gamma]^\prime$ that minimizes equation~\eqref{eq3.6} at each iteration, for a total of $N_{\text{max}}=300$ iterations in this study. An optimal set of parameters is obtained once the value of the objective function $J$ converges, meaning that the difference of $J$ from one iteration to the next is sufficiently small, as defined by equation~\eqref{eq4.5}. For illustrative purposes, Figure~\ref{Objective_ScenarioI} shows the objective function value for MPR = 10\%, corresponding to the case of a single AV in the middle of the platoon, with a step size of $\epsilon=1\times10^{-5}$. The value of $J$ is observed to decrease as the number of iterations increases, indicating convergence of the computation. Further, we calculate the ASV and average FC for all the 10 vehicles following the leader, for various feasible values of $\theta$ under 10\% MPR, with the corresponding results shown in Figure~\ref{ASV_FC_Unoptimal_ScenarioI_10MPR}. Values for both ASV (Figure~\ref{ASV_Unoptimal_ScenarioI_10MPR}) and FC (Figure~\ref{FC_Unoptimal_ScenarioI_10MPR}) are observed to decrease with higher values of $\beta$ and $\gamma$, gradually approaching the optimum, indicating the superior performance of the optimal control parameter values compared to their non-optimal counterparts. Following expression~\eqref{eq3.5}, the upper bound of $\beta$ is determined as $2[s_{i}(0)-\check{s}]/(\pi t_{\textnormal{f}}) = 2(52.42-2)/(500\pi) = 0.0642$, where the minimum safe spacing $\check{s}$ is set the same as $s_0$ of the IDM~\citep{wang2023general}. Additionally, the impact of $\beta$ on the decrease of both ASV and FC appears to be greater than that of $\gamma$, indicated by the steeper slope along the $\beta$-axis. This is consistent with the analytical form of the additive controller, TS-OPS, given by equation~\eqref{eq3.2}, whose effectiveness is directly affected by the value of $\beta$, as opposed to the effect of $\gamma$ being limited by the natural bounds of the $\arctan(\cdot)$ function.

 \begin{figure}[t!]
    \centering
    \includegraphics[width=0.6\linewidth]{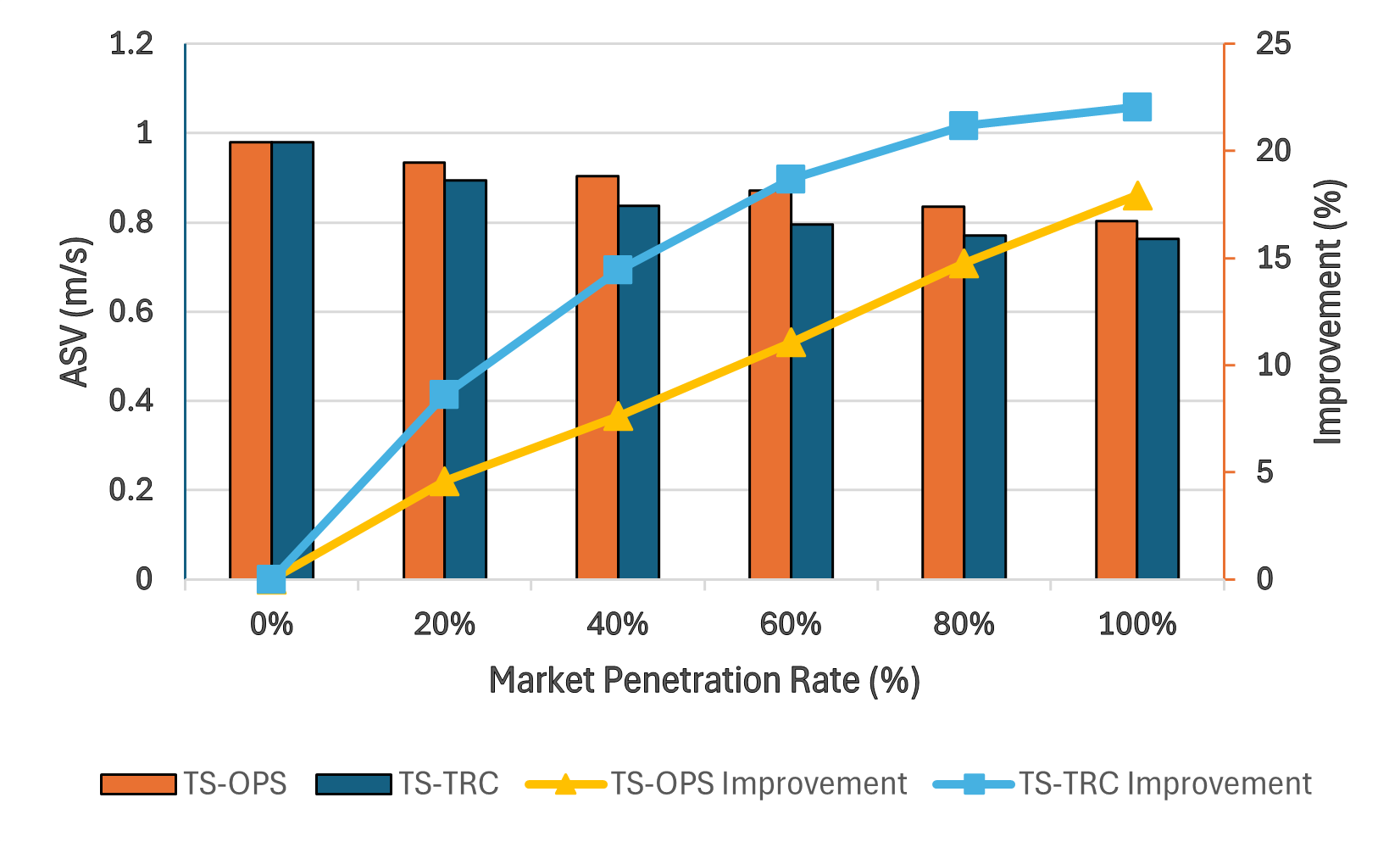}
    \vskip-12pt
    \caption{Scenario~I: Analysis of ASV under multiple MPRs with AVs using TS-OPS and TS-TRC. A greater presence of AVs using the TS-OPS leads to higher reductions in traffic oscillations, effectively smoothing the flow. Although the TS-TRC, developed in our recent work~\citep{wang2023general}, achieves better results than the proposed TS-OPS, it relies on the strong assumption of a known equilibrium traffic speed, which is typically challenging to obtain in the real world. The TS-OPS performs well in reducing ASV, while offering much higher implementability.}
    \label{ASV_ScenarioI}
\end{figure}

 \begin{figure}[t!]
    \centering
    \includegraphics[width=0.6\linewidth]{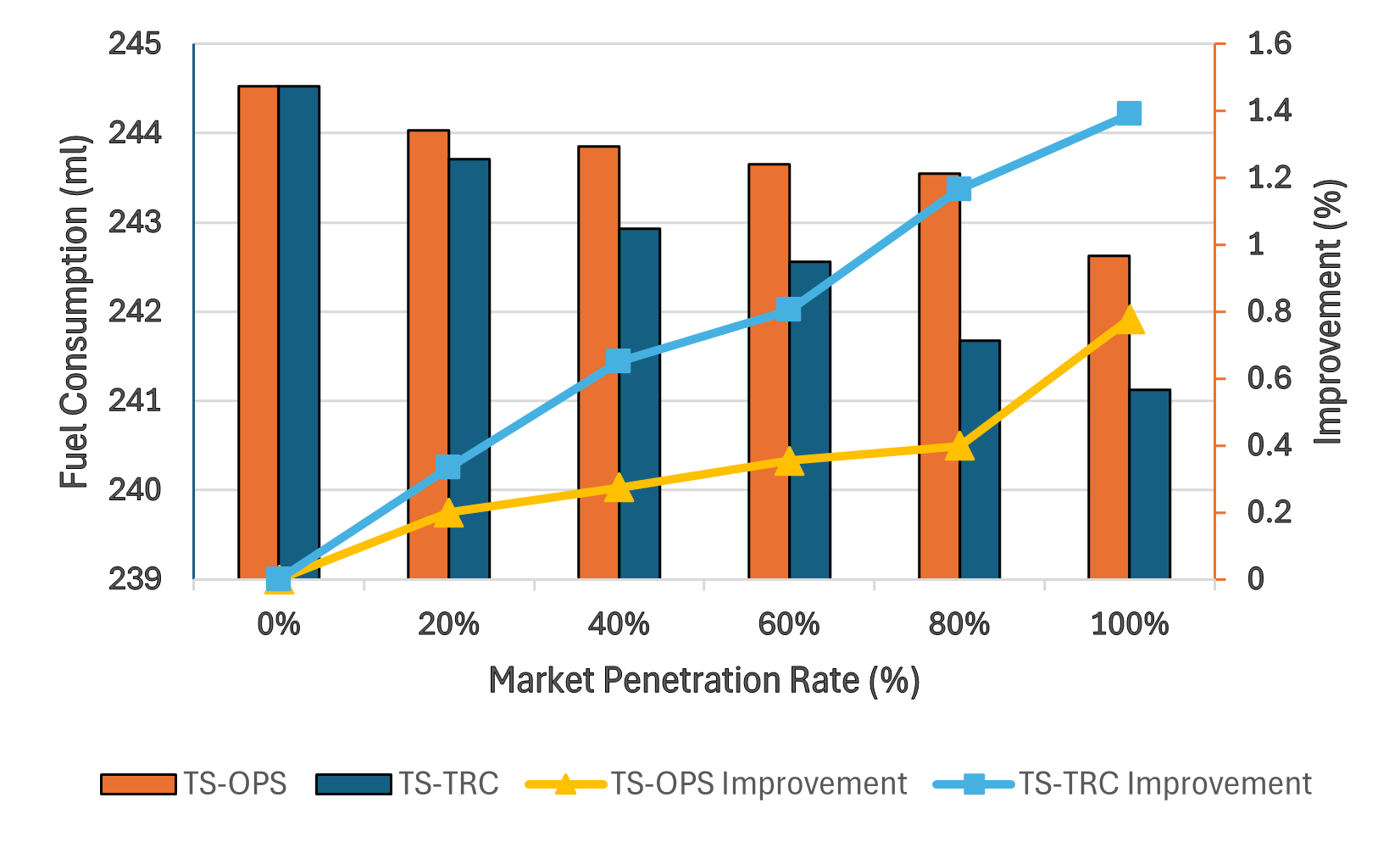}
    \vskip-12pt
    \caption{Scenario~I: Analysis of FC under multiple MPRs with AVs using TS-OPS and TS-TRC. Since a higher MPR of AVs leads to smoother traffic flow, a greater reduction in FC is observed as more AVs using the proposed additive controller TS-OPS are introduced into the platoon. Although the TS-OPS does not perform as well as the latest AV additive controller TS-TRC, which relies on impractical time-varying traffic information, it is still effective in reducing vehicle energy consumption and improving traffic efficiency.} 
    \label{FC_ScenarioI}
\end{figure}

Next, we present comprehensive results to measure and quantify the effectiveness of the proposed AV control approach in smoothing traffic and improving efficiency. These results are summarized in Figures~\ref{ASV_ScenarioI} and~\ref{FC_ScenarioI}, showing the ASV and average FC of the platoon over a range of AV MPRs. Figure~\ref{ASV_ScenarioI} shows that the ASV values corresponding to the proposed TS-OPS controller decrease with the increase of AV MPR, indicating smoother flow with fewer oscillations as more intelligently controlled AVs are introduced into the traffic. Consequently, ASV improves by up to 18\% under full AV penetration. A smoother traffic flow leads to reduced fuel consumption, as shown in Figure~\ref{FC_ScenarioI}, where a similar trend is observed for the average FC. It decreases from an initial value of 244.52~ml to 242.62~ml under 100\% MPR due to the reduction of stop-and-go traffic waves. These results are consistent with the recent findings of~\citep{sun2022energy}, where AVs are assumed to communicate with each other within a certain range (which is not required in this study).

In addition to optimally selecting the control parameters $\beta$ and $\gamma$, a distinctive feature of the proposed TS-OPS controller is that it does not require knowledge of the equilibrium traffic speed, which is challenging to obtain. This is in contrast to the theoretical additive controller, TS-TRC, developed in our recent study~\citep{wang2023general}. As a result, the TS-OPS is more implementable, requiring only local traffic information $s_i$ and $\Delta v_i$, which are readily accessible through the onboard sensors of an AV. We compare the performance of the TS-OPS with the TS-TRC in terms of ASV and average FC in Figures~\ref{ASV_ScenarioI} and~\ref{FC_ScenarioI}, respectively. Following~\citep{wang2023general}, the additive AV controller, TS-TRC, is defined as $u_i = \Delta v_i + 0.1 \cdot \arctan[0.01 s_i (v^* - v_{i-1})]$ with model parameter values shown in Table~\ref{Table_parameters_ScenarioI}. The TS-TRC slightly outperforms the TS-OPS in both metrics due to its assumed knowledge of the equilibrium speed $v^*$, typically unavailable to vehicles. The TS-TRC reduces ASV by up to 22\% compared to 18\% for the TS-OPS and lowers the average FC by up to 1.39\%, 0.61\% more than the reduction achieved by the TS-OPS. Overall, the TS-TRC achieves slightly better results than the TS-OPS due to its additional information about the desirable traffic state. However, the TS-OPS still performs well in reducing traffic oscillations and improving efficiency, while offering higher implementability.

\subsubsection{Scenario II}\label{section4.3.2}

\begin{figure}[t!]
 	\centering
 	\subfloat[\textnormal{MPR = 0\%}]{\includegraphics[width=0.33\textwidth]{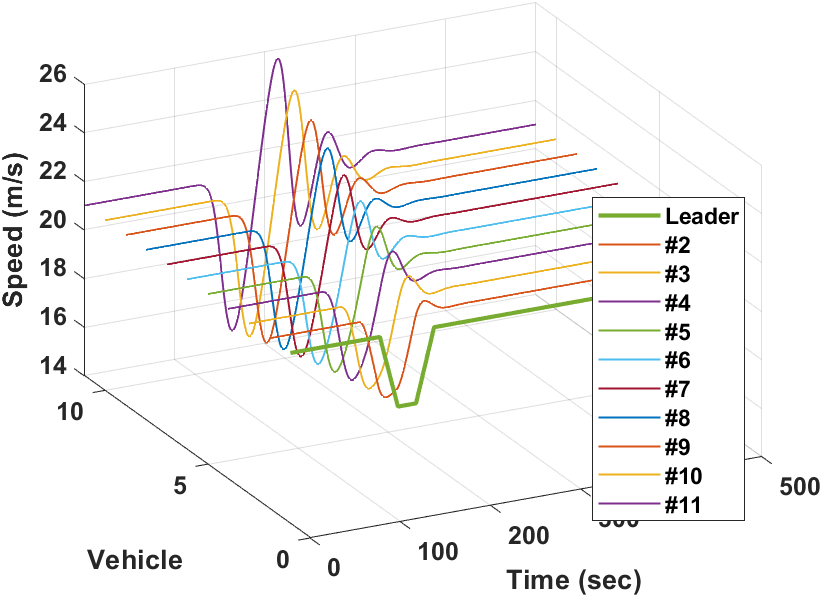}\label{Speed_0_Scenario2}}
 	\hfil%
 	\subfloat[\textnormal{MPR = 50\%}]{\includegraphics[width=0.33\textwidth]{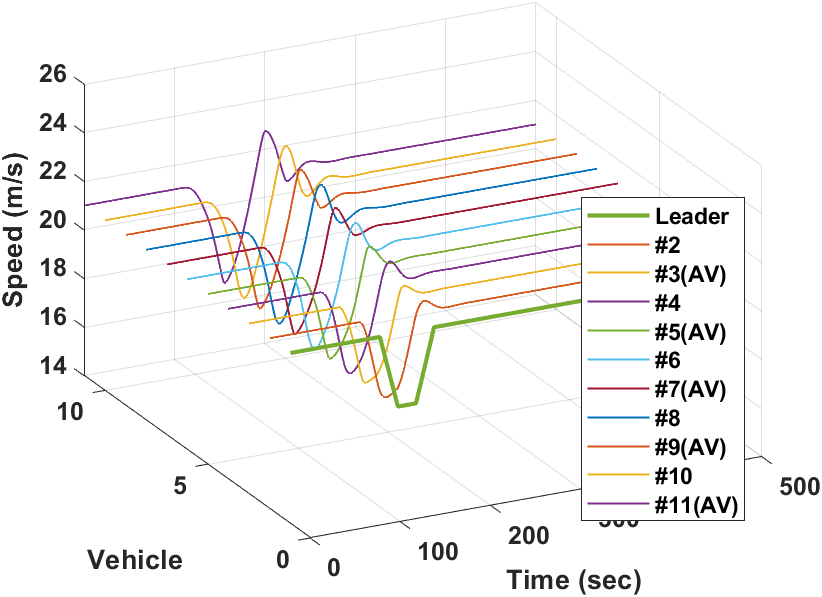}\label{Speed_50_Scenario2}}
 	\hfil%
 	\subfloat[\textnormal{MPR = 100\%}]{\includegraphics[width=0.33\textwidth]{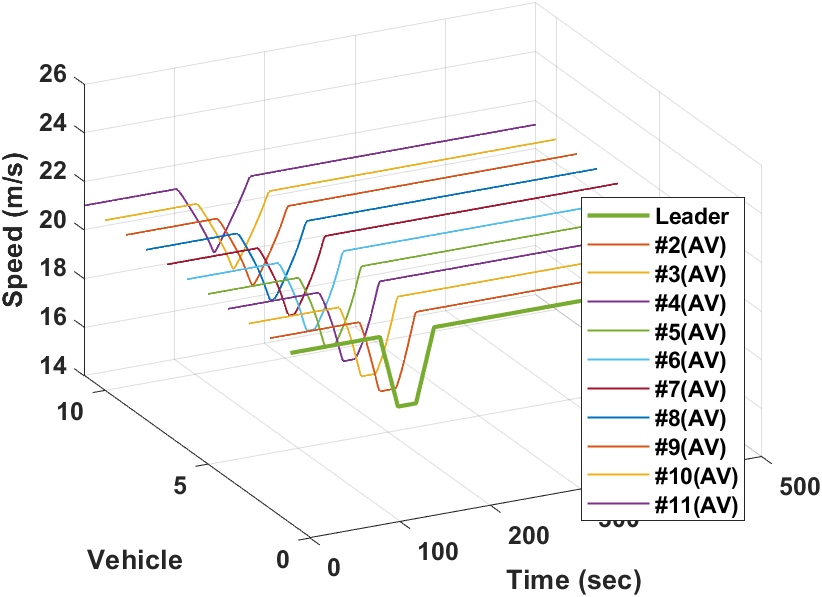}\label{Speed_100_Scenario2}}
    \caption{Scenario~II: 3D plot of vehicle speeds under MPR = 0\%, 50\%, and 100\%. AVs are controlled with the TS-OPS developed in this study, while HVs follow the IDM with parameter values shown in Table~\ref{Table_parameters_ScenarioII}. As the MPR of AVs increases, stop-and-go traffic waves decrease, leading to smoother traffic flow.}\label{Speed_ScenarioII}
\end{figure}

\begin{figure}[t!]
 	\centering
 	\subfloat[\textnormal{MPR = 0\%}]{\includegraphics[width=0.33\textwidth]{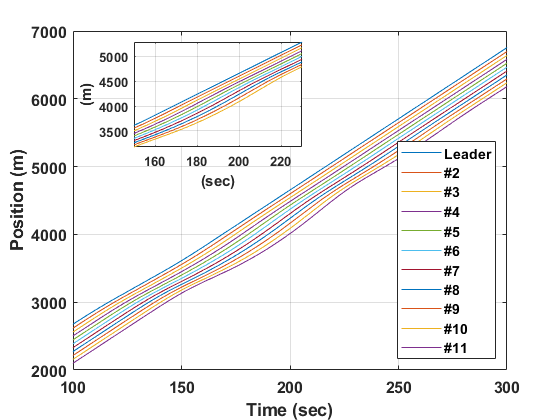}\label{Position_0_Scenario2}}
 	\hfil%
 	\subfloat[\textnormal{MPR = 50\%}]{\includegraphics[width=0.33\textwidth]{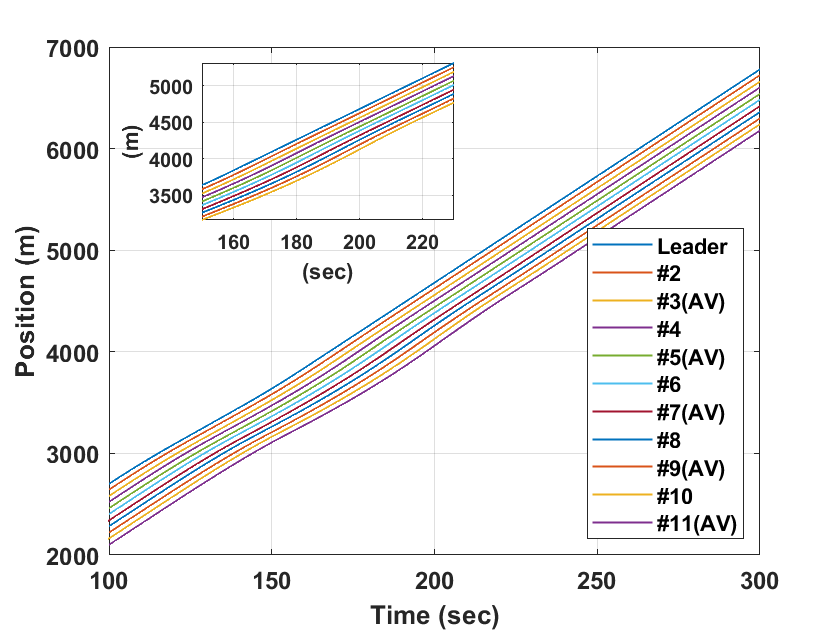}\label{Position_50_Scenario2}}
 	\hfil%
 	\subfloat[\textnormal{MPR = 100\%}]{\includegraphics[width=0.33\textwidth]{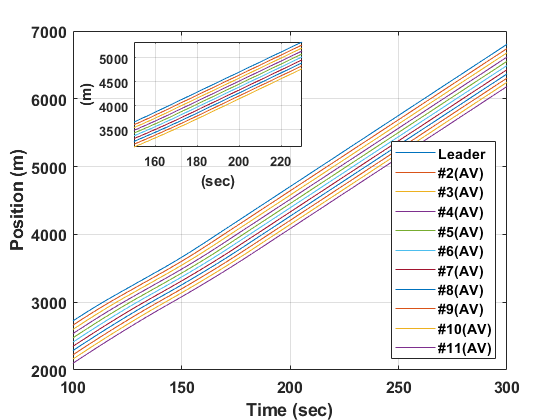}\label{Position_100_Scenario2}}
    \caption{Scenario~II: 2D plot of vehicle trajectories (positions) under MPR = 0\%, 50\%, and 100\%, corresponding to the speed profiles presented in Figure~\ref{Speed_ScenarioII}. Traffic becomes smoother as more AVs are introduced into the platoon, and car-following safety is ensured with no rear-end collisions observed.}
    \label{Safety_ScenarioII}
\end{figure}

This scenario adopts a different set of parameter values for the IDM, as shown in Table~\ref{Table_parameters_ScenarioII}. These values are calibrated based on real-world car-following experiments described in~\citep{de2021calibrating}, representing highly unstable driving behavior. They are used here to assess the effectiveness of the TS-OPS in smoothing traffic with greater levels of oscillation, which allows for a clear demonstration of its efficacy~\citep{wang2023general,aguilar2024energy}. Figure~\ref{Speed_0_Scenario2} shows the speed profiles of all vehicles in the absence of AVs, where the slowdown of the leader amplifies and propagates upstream, causing stop-and-go traffic waves. This set of IDM parameter values results in greater speed oscillations compared to those used in Scenario~I, as seen from the comparison between Figure~\ref{Speed_0_Scenario2} and Figure~\ref{Speed_0_Scenario1}. As AVs with the TS-OPS controller are introduced into the platoon, vehicle speed oscillations decrease significantly, leading to shorter periods of perturbation and smoother traffic flow, as observed in Figures~\ref{Speed_50_Scenario2} and~\ref{Speed_100_Scenario2} for 50\% and 100\% MPRs, respectively. The corresponding vehicle trajectories are presented in Figure~\ref{Safety_ScenarioII}, showing no risks of rear-end collisions, which indicates car-following safety for AVs using the TS-OPS controller. To capture the totality of the oscillations in the simulation, we consider the period from 100 to 300 seconds when calculating ASV and FC.

\begin{table}[t!]
\setlength{\tabcolsep}{0.8pt}
\caption{Model parameter values of the IDM~\citep{de2021calibrating,wang2023general}}\label{Table_parameters_ScenarioII}
\vspace*{-2mm}
\begin{center}
 \begin{tabular}{c c c c c c c c}
 \hline
 \textbf{IDM} ~&~ $a$ (m/s\textsuperscript{2}) ~&~ $b$ (m/s\textsuperscript{2}) ~&~ $v_{0}$ (m/s) ~&~ $s_{0}$ (m) ~&~ $T$ (s) ~&~ $\delta$ ~&~ $l_{i}$ (m)  \\  [0.5ex]
 \textbf{Value} ~&~ 0.6 ~&~ 5.2 ~&~ 44.1 ~&~ 6.3 ~&~ 2.2 ~&~ 15.5 ~&~ 5 \\ [0.3ex]
 \hline
\end{tabular}
\end{center}
\end{table}

\begin{figure}[t!]
    \centering
    \includegraphics[width=0.4\linewidth]{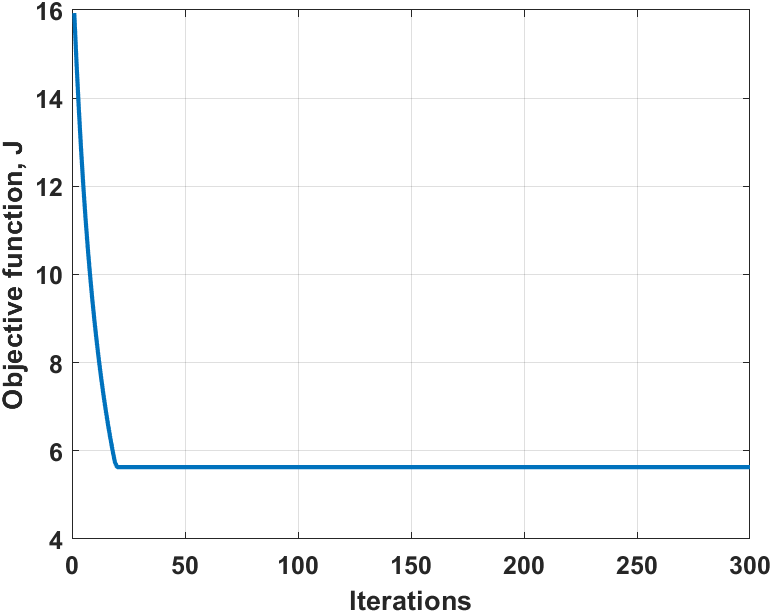}
    \vskip-6pt
    \caption{Scenario~II: The objective function $J$ converges as the number of iterations increases under a 10\% MPR.}
    \label{Objective_ScenarioII}
\end{figure}

\begin{figure}[t!]
    \centering
    \subfloat[\textnormal{ASV results at 10\% MPR}]{\includegraphics[width=0.4\linewidth]{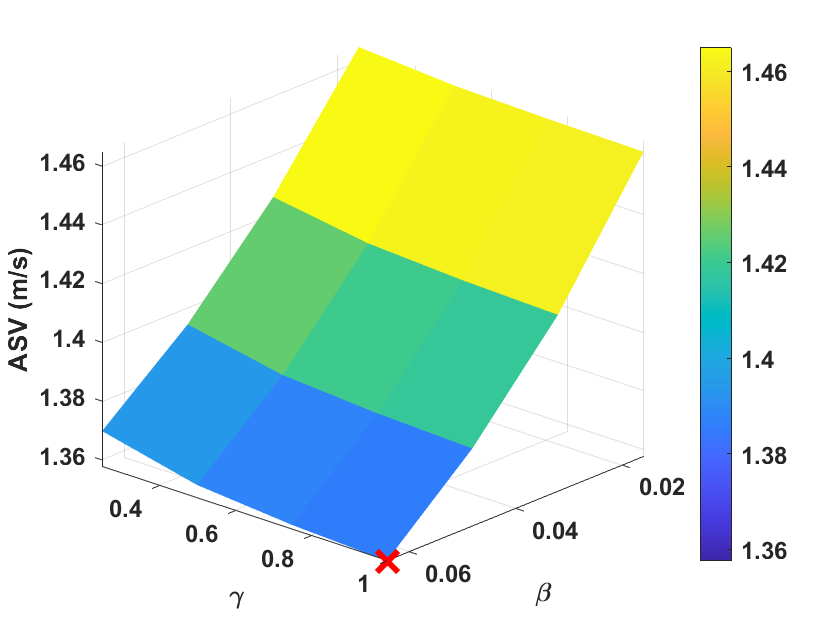}\label{ASV_Unoptimal_ScenarioII_10MPR}}
    \hfil%
    \subfloat[\textnormal{FC results at 10\% MPR}]{\includegraphics[width=0.4\linewidth]{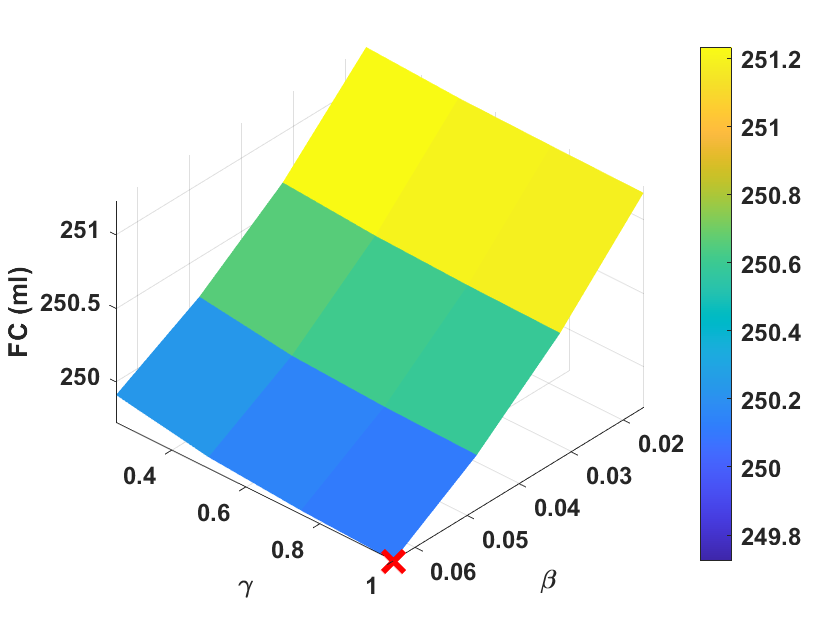}\label{FC_Unoptimal_ScenarioII_10MPR}}
    \caption{Scenario II: ASV and average FC results at 10\% MPR under multiple combinations of non-optimal $\beta$ and $\gamma$. The red marker represents the optimal values of ASV and FC in Figures~\ref{ASV_Unoptimal_ScenarioII_10MPR} and~\ref{FC_Unoptimal_ScenarioII_10MPR}, respectively, obtained at the optimal control parameters, $\beta = 0.0642$ and $\gamma = 1.0017$. The optimal value of $\beta$ reaches its upper bound as determined by expression~\eqref{eq3.5}.}
    \label{ASV_FC_Unoptimal_ScenarioII_10MPR}
\end{figure}

For illustration, the value of the objective function $J$ under 10\% MPR is shown in Figure~\ref{Objective_ScenarioII}, which is observed to decrease as the iteration number increases, indicating computational convergence. As seen in Scenario~I, we also calculate the ASV and average FC for the 10-vehicle platoon, for various non-optimal values of $\theta$ under 10\% MPR. The corresponding results are shown in Figure~\ref{ASV_FC_Unoptimal_ScenarioII_10MPR}. Values for both ASV (Figure~\ref{ASV_Unoptimal_ScenarioII_10MPR}) and FC (Figure~\ref{FC_Unoptimal_ScenarioII_10MPR}) are observed to decrease with higher values of $\beta$ and $\gamma$ toward the optimum (corresponding to optimal control parameters) within 300 iterations. This is similar to the results presented in Figure~\ref{ASV_FC_Unoptimal_ScenarioI_10MPR} for Scenario~I. 

\begin{figure}[t!]
    \centering
    \includegraphics[width=0.6\linewidth]{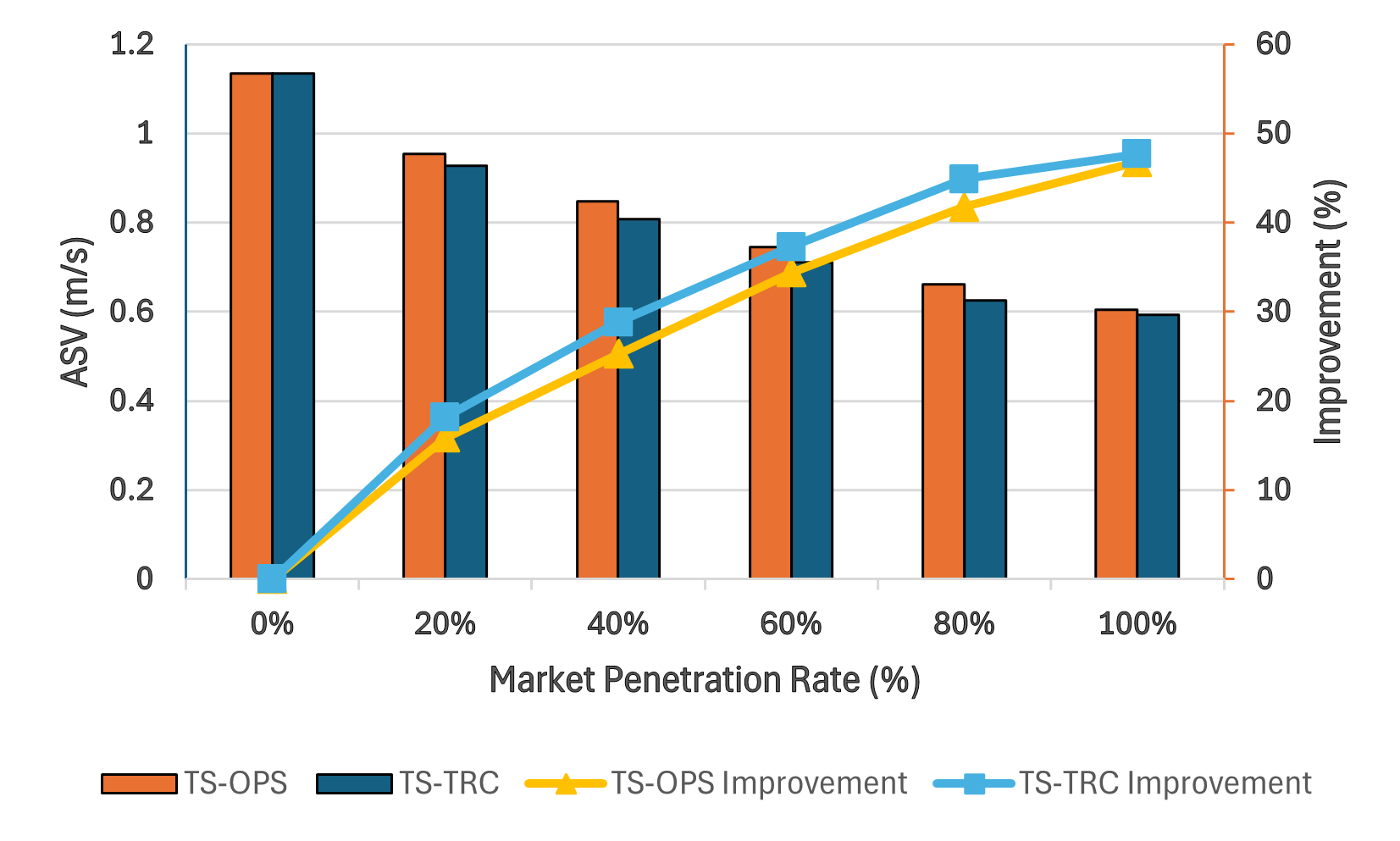}
    \vskip-12pt
    \caption{Scenario~II: ASV results under multiple MPRs for TS-OPS and TS-TRC. The newly proposed TS-OPS is highly effective in reducing ASV and smoothing the flow, even in highly unstable and oscillatory traffic conditions. It performs nearly as well as the recent TS-TRC developed in~\citep{wang2023general}.} 
    \label{ASV_ScenarioII}
\end{figure}

\begin{figure}[t!]
    \centering
    \includegraphics[width=0.6\linewidth]{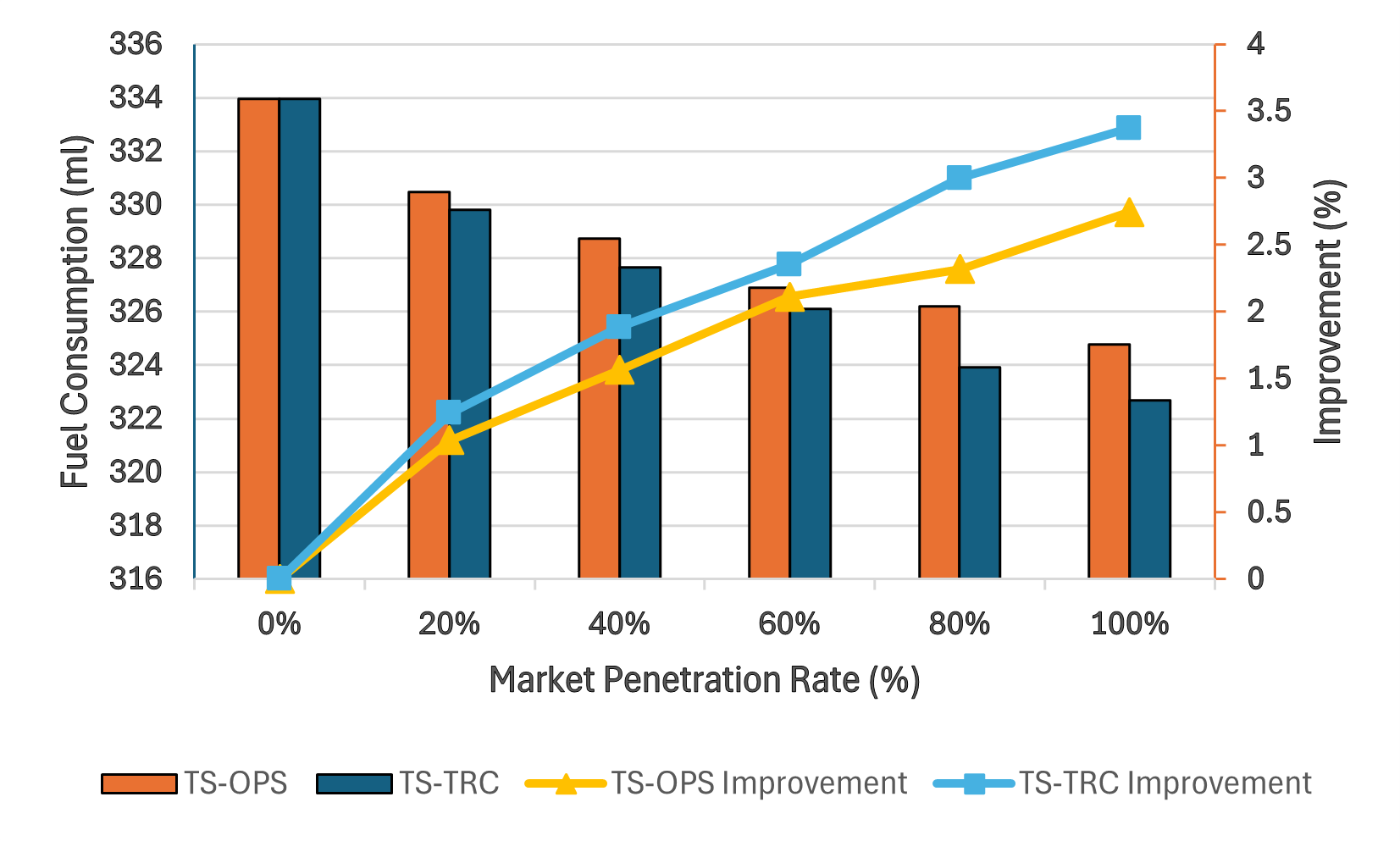}
    \vskip-12pt
    \caption{Scenario~II: FC results under multiple MPRs for TS-OPS and TS-TRC. Due to its traffic-smoothing capability, the proposed TS-OPS is effective in lowering energy consumption for the platoon, despite highly oscillatory traffic conditions. Its performance is slightly inferior compared to that of TS-TRC.}
    \label{FC_ScenarioII}
\end{figure}

Similar to Scenario~I, we present a set of results to quantify the effectiveness of the proposed TS-OPS controller in smoothing traffic and improving efficiency, as shown in Figures~\ref{ASV_ScenarioII} and~\ref{FC_ScenarioII} for a range of MPRs. Figure~\ref{ASV_ScenarioII} shows that AVs controlled with the TS-OPS reduce ASV across all the MPRs considered, with higher reductions observed at a greater presence of AVs. This leads to fewer traffic oscillations and smoother traffic flow, thereby resulting in less fuel consumption from vehicles, as illustrated in Figure~\ref{FC_ScenarioII}. Specifically, ASV achieves an improvement of up to 46.78\% when all vehicles are automated (MPR=100\%) using the proposed TS-OPS, accompanied by a 2.74\% reduction in FC across the platoon. Further, in terms of both ASV and FC, the proposed TS-OPS slightly underperforms the additive AV controller TS-TRC ($u_i = \Delta v_i + 0.04 \cdot \arctan[0.01 s_i (v^* - v_{i-1})]$) developed in our recent study~\citep{wang2023general}, as shown in Figures~\ref{ASV_ScenarioII} and~\ref{FC_ScenarioII}. Under 100\% MPR, the TS-TRC achieves an improvement of 47.69\% and 3.37\% in ASV and FC, respectively, exceeding their respective counterparts for the TS-OPS by only 0.91\% and 0.63\%. Overall, the proposed TS-OPS controller performs well, considering such slight differences in performance improvement compared to the latest additive AV controller TS-TRC. As mentioned in Scenario~I, the slightly superior performance of the TS-TRC is attributed to its strict assumption of the knowledge of the equilibrium traffic speed, rendering its practical applicability limited.

\begin{remark}\label{Remark5}
As discussed in Remark~\ref{Remark4}, the upper bound on $\beta$ given by~\eqref{eq3.5} is derived conservatively in that an AV may not experience perturbations throughout the entire time horizon $I$. Consequently, the duration of traffic waves experienced by the vehicle is likely shorter than $I$. Therefore, the performance of the TS-OPS may be further improved by using a larger upper bound on $\beta$, depending on the actual period of speed perturbation (at most $t_{\textnormal{f}}$).
\end{remark}

\section{Conclusions}\label{section5}

The development of properly controlled automated vehicles (AVs) will allow their use as mobile actuators, making it possible for them to alleviate stop-and-go traffic waves caused by the collective behavior of human drivers. In this study, we have proposed a new additive AV controller, TS-OPS, which is shown to be effective in smoothing traffic flow, thereby reducing vehicle energy consumption and improving traffic efficiency. A distinct feature of the TS-OPS is that its synthesis requires only local traffic information, such as inter-vehicle spacing and relative speed, which are readily available through AV onboard sensors. Essentially, it allows a controlled AV to track a subtler version of the perturbed speed profile resulting from its preceding vehicle, thereby enabling smoother traffic flow. Additionally, we provide a method for selecting the optimal control parameters to achieve traffic-smoothing effects efficiently.

Two distinct traffic scenarios with varying levels of oscillation have been considered to examine the effectiveness of the proposed approach. Numerical results show that AVs using the TS-OPS are capable of reducing traffic oscillations and lowering vehicle fuel consumption, thereby improving traffic performance. The traffic-smoothing effect of the TS-OPS is more pronounced at higher penetration rates of AVs. Although the performance of the proposed TS-OPS is slightly less superior to the latest additive AV controller, TS-TRC, developed in our recent study, it eliminates the strict requirement of equilibrium traffic speed, which is necessary for synthesizing TS-TRC and challenging to obtain in the real world.

While this study has developed an effective approach for directly smoothing nonlinear traffic through the intelligent control of AVs, further research efforts could explore its holistic influence on the transportation system in more realistic traffic scenarios. Additionally, it would be of significant interest to extend the TS-OPS controller to incorporate its impact on HVs in the design process, motivated by the ideas presented in~\citep{shang2024interaction}.

\bibliographystyle{unsrtnat}
\bibliography{mybibfile} 

\begin{thebibliography}{39}
\providecommand{\natexlab}[1]{#1}
\providecommand{\url}[1]{\texttt{#1}}
\expandafter\ifx\csname urlstyle\endcsname\relax
  \providecommand{\doi}[1]{doi: #1}\else
  \providecommand{\doi}{doi: \begingroup \urlstyle{rm}\Url}\fi

\bibitem[Laval and Daganzo(2006)]{laval2006lane}
Jorge~A Laval and Carlos~F Daganzo.
\newblock Lane-changing in traffic streams.
\newblock \emph{Transportation Research Part B: Methodological}, 40\penalty0
  (3):\penalty0 251--264, 2006.

\bibitem[Kerner(2012)]{kerner2012physics}
Boris~S Kerner.
\newblock \emph{The physics of traffic: Empirical freeway pattern features,
  engineering applications, and theory}.
\newblock Springer, 2012.

\bibitem[Milan{\'e}s et~al.(2010)Milan{\'e}s, Godoy, Villagr{\'a}, and
  P{\'e}rez]{milanes2010automated}
Vicente Milan{\'e}s, Jorge Godoy, Jorge Villagr{\'a}, and Joshu{\'e} P{\'e}rez.
\newblock Automated on-ramp merging system for congested traffic situations.
\newblock \emph{IEEE Transactions on Intelligent Transportation Systems},
  12\penalty0 (2):\penalty0 500--508, 2010.

\bibitem[Flynn et~al.(2009)Flynn, Kasimov, Nave, Rosales, and
  Seibold]{flynn2009self}
Morris~R Flynn, Aslan~R Kasimov, J-C Nave, Rodolfo~R Rosales, and Benjamin
  Seibold.
\newblock Self-sustained nonlinear waves in traffic flow.
\newblock \emph{Physical Review E}, 79\penalty0 (5):\penalty0 056113, 2009.

\bibitem[Sugiyama et~al.(2008)]{sugiyama2008traffic}
Yuki Sugiyama et~al.
\newblock Traffic jams without bottlenecks—experimental evidence for the
  physical mechanism of the formation of a jam.
\newblock \emph{New Journal of Physics}, 10\penalty0 (3):\penalty0 033001,
  2008.

\bibitem[Wu et~al.(2019)]{wu2019tracking}
Fangyu Wu et~al.
\newblock Tracking vehicle trajectories and fuel rates in phantom traffic jams:
  Methodology and data.
\newblock \emph{Transportation Research Part C: Emerging Technologies},
  99:\penalty0 82--109, 2019.

\bibitem[Wang et~al.(2019)Wang, Ahmed, and Yeap]{ahmed2018optimum}
Shian Wang, Nasir~Uddin Ahmed, and Tet~H Yeap.
\newblock Optimum management of urban traffic flow based on a stochastic
  dynamic model.
\newblock \emph{IEEE Transactions on Intelligent Transportation Systems},
  20\penalty0 (12):\penalty0 4377--4389, 2019.

\bibitem[Shang et~al.(2023)Shang, Wang, and Stern]{shang2023extending}
Mingfeng Shang, Shian Wang, and Raphael~E Stern.
\newblock Extending ramp metering control to mixed autonomy traffic flow with
  varying degrees of automation.
\newblock \emph{Transportation Research Part C: Emerging Technologies},
  151:\penalty0 104119, 2023.

\bibitem[Smulders(1990)]{smulders1990control}
Stef Smulders.
\newblock Control of freeway traffic flow by variable speed signs.
\newblock \emph{Transportation Research Part B: Methodological}, 24\penalty0
  (2):\penalty0 111--132, 1990.

\bibitem[Bayen et~al.(2022)Bayen, Delle~Monache, Garavello, Goatin, and
  Piccoli]{bayen2022control}
Alexandre Bayen, Maria~Laura Delle~Monache, Mauro Garavello, Paola Goatin, and
  Benedetto Piccoli.
\newblock Control problems for conservation laws with traffic applications:
  Modeling, analysis, and numerical methods, 2022.

\bibitem[Zheng et~al.(2020)Zheng, Wang, and Li]{zheng2020smoothing}
Yang Zheng, Jiawei Wang, and Keqiang Li.
\newblock Smoothing traffic flow via control of autonomous vehicles.
\newblock \emph{IEEE Internet of Things Journal}, 7\penalty0 (5):\penalty0
  3882--3896, 2020.

\bibitem[Zhu and Zhang(2018)]{zhu2018analysis}
Wen~Xing Zhu and H~M Zhang.
\newblock Analysis of mixed traffic flow with human-driving and autonomous cars
  based on car-following model.
\newblock \emph{Physica A: Statistical Mechanics and its Applications},
  496:\penalty0 274--285, 2018.

\bibitem[Wang et~al.(2022{\natexlab{a}})Wang, Stern, and
  Levin]{wang2022optimal}
Shian Wang, Raphael Stern, and Michael~W Levin.
\newblock Optimal control of autonomous vehicles for traffic smoothing.
\newblock \emph{IEEE Transactions on Intelligent Transportation Systems},
  23\penalty0 (4):\penalty0 3842--3852, 2022{\natexlab{a}}.

\bibitem[Wang et~al.(2023{\natexlab{a}})Wang, Shang, Levin, and
  Stern]{wang2023general}
Shian Wang, Mingfeng Shang, Michael~W Levin, and Raphael Stern.
\newblock A general approach to smoothing nonlinear mixed traffic via control
  of autonomous vehicles.
\newblock \emph{Transportation Research Part C: Emerging Technologies},
  146:\penalty0 103967, 2023{\natexlab{a}}.

\bibitem[Lichtl{\'e} et~al.(2022)Lichtl{\'e}, Vinitsky, Nice, Seibold, Work,
  and Bayen]{lichtle2022deploying}
Nathan Lichtl{\'e}, Eugene Vinitsky, Matthew Nice, Benjamin Seibold, Dan Work,
  and Alexandre~M Bayen.
\newblock Deploying traffic smoothing cruise controllers learned from
  trajectory data.
\newblock In \emph{2022 International Conference on Robotics and Automation
  (ICRA)}, pages 2884--2890. IEEE, 2022.

\bibitem[Lichtl{\'e} et~al.(2023)Lichtl{\'e}, Jang, Shah, Vinitsky, Lee, and
  Bayen]{lichtle2023traffic}
Nathan Lichtl{\'e}, Kathy Jang, Adit Shah, Eugene Vinitsky, Jonathan~W Lee, and
  Alexandre~M Bayen.
\newblock Traffic smoothing controllers for autonomous vehicles using deep
  reinforcement learning and real-world trajectory data.
\newblock In \emph{2023 IEEE 26th International Conference on Intelligent
  Transportation Systems (ITSC)}, pages 4346--4351. IEEE, 2023.

\bibitem[Li et~al.(2024)Li, Wang, Shang, Choi, and Stern]{li2024customizable}
Tianyi Li, Shian Wang, Mingfeng Shang, Seongjin Choi, and Raphael Stern.
\newblock A customizable neural network based framework for autonomous vehicle
  control with human-guided learning.
\newblock In \emph{2024 IEEE 27th International Conference on Intelligent
  Transportation Systems (ITSC)}. IEEE, 2024.

\bibitem[Wang et~al.(2023{\natexlab{b}})Wang, Lian, Jiang, Xu, Li, and
  Jones]{wang2023distributed}
Jiawei Wang, Yingzhao Lian, Yuning Jiang, Qing Xu, Keqiang Li, and Colin~N
  Jones.
\newblock Distributed data-driven predictive control for cooperatively
  smoothing mixed traffic flow.
\newblock \emph{Transportation Research Part C: Emerging Technologies},
  155:\penalty0 104274, 2023{\natexlab{b}}.

\bibitem[Wang et~al.(2022{\natexlab{b}})Wang, Li, and Levin]{wang2022policy}
Shian Wang, Zhexian Li, and Michael~W Levin.
\newblock Optimal policy for integrating autonomous vehicles into the auto
  market.
\newblock \emph{Transportation Research Part C: Emerging Technologies},
  143:\penalty0 103821, 2022{\natexlab{b}}.

\bibitem[Wilson and Ward(2011)]{wilson2011car}
R~Eddie Wilson and Jonathan~A Ward.
\newblock Car--following models: Fifty years of linear stability analysis--a
  mathematical perspective.
\newblock \emph{Transportation Planning and Technology}, 34\penalty0
  (1):\penalty0 3--18, 2011.

\bibitem[Cui et~al.(2017)Cui, Seibold, Stern, and Work]{cui2017stabilizing}
Shumo Cui, Benjamin Seibold, Raphael Stern, and Daniel~B Work.
\newblock Stabilizing traffic flow via a single autonomous vehicle:
  Possibilities and limitations.
\newblock In \emph{2017 IEEE Intelligent Vehicles Symposium (IV)}, pages
  1336--1341. IEEE, 2017.

\bibitem[Wu et~al.(2018)Wu, Bayen, and Mehta]{wu2018stabilizing}
Cathy Wu, Alexandre~M Bayen, and Ankur Mehta.
\newblock Stabilizing traffic with autonomous vehicles.
\newblock In \emph{2018 IEEE International Conference on Robotics and
  Automation (ICRA)}, pages 6012--6018. IEEE, 2018.

\bibitem[Stern et~al.(2018)]{stern2018dissipation}
Raphael~E Stern et~al.
\newblock Dissipation of stop-and-go waves via control of autonomous vehicles:
  Field experiments.
\newblock \emph{Transportation Research Part C: Emerging Technologies},
  89:\penalty0 205--221, 2018.

\bibitem[Giammarino et~al.(2021)Giammarino, Baldi, Frasca, and
  Delle~Monache]{giammarino2020traffic}
Vittorio Giammarino, Simone Baldi, Paolo Frasca, and Maria~Laura Delle~Monache.
\newblock Traffic flow on a ring with a single autonomous vehicle: An
  interconnected stability perspective.
\newblock \emph{IEEE Transactions on Intelligent Transportation Systems},
  22\penalty0 (8):\penalty0 4998--5008, 2021.

\bibitem[Wang et~al.(2022{\natexlab{c}})Wang, Shang, Levin, and
  Stern]{wang2022smoothing}
Shian Wang, Mingfeng Shang, Michael~W Levin, and Raphael Stern.
\newblock Smoothing nonlinear mixed traffic with autonomous vehicles: Control
  design.
\newblock In \emph{2022 IEEE 25th International Conference on Intelligent
  Transportation Systems (ITSC)}, pages 661--666. IEEE, 2022{\natexlab{c}}.

\bibitem[Delle~Monache et~al.(2022)Delle~Monache, Pasquale, Barreau, and
  Stern]{delle2022new}
M~L Delle~Monache, C~Pasquale, Matthieu Barreau, and R~Stern.
\newblock New frontiers of freeway traffic control and estimation.
\newblock In \emph{2022 IEEE 61st Conference on Decision and Control (CDC)},
  pages 6910--6925. IEEE, 2022.

\bibitem[Ahmed(1976)]{ahmed1976simple}
Nasir~Uddin Ahmed.
\newblock A simple gradient algorithm for least squares estimation of system
  parameters.
\newblock \emph{International Journal of Systems Science}, 7\penalty0
  (6):\penalty0 673--677, 1976.

\bibitem[Treiber et~al.(2000)Treiber, Hennecke, and
  Helbing]{treiber2000congested}
Martin Treiber, Ansgar Hennecke, and Dirk Helbing.
\newblock Congested traffic states in empirical observations and microscopic
  simulations.
\newblock \emph{Physical Review E}, 62:\penalty0 1805, 2000.

\bibitem[Treiber and Kesting(2013)]{treiber2013traffic}
Martin Treiber and Arne Kesting.
\newblock Traffic flow dynamics.
\newblock \emph{Traffic Flow Dynamics: Data, Models and Simulation,
  Springer-Verlag Berlin Heidelberg}, pages 983--1000, 2013.

\bibitem[Gunter et~al.(2019)Gunter, Stern, and Work]{gunter2019modeling}
George Gunter, Raphael Stern, and Daniel~B Work.
\newblock Modeling adaptive cruise control vehicles from experimental data:
  model comparison.
\newblock In \emph{2019 IEEE Intelligent Transportation Systems Conference
  (ITSC)}, pages 3049--3054. IEEE, 2019.

\bibitem[Milan{\'e}s et~al.(2013)Milan{\'e}s, Shladover, Spring, Nowakowski,
  Kawazoe, and Nakamura]{milanes2013cooperative}
Vicente Milan{\'e}s, Steven~E Shladover, John Spring, Christopher Nowakowski,
  Hiroshi Kawazoe, and Masahide Nakamura.
\newblock Cooperative adaptive cruise control in real traffic situations.
\newblock \emph{IEEE Transactions on Intelligent Transportation Systems},
  15\penalty0 (1):\penalty0 296--305, 2013.

\bibitem[Liang and Peng(1999)]{liang1999optimal}
Chi-Ying Liang and Huei Peng.
\newblock Optimal adaptive cruise control with guaranteed string stability.
\newblock \emph{Vehicle System Dynamics}, 32\penalty0 (4-5):\penalty0 313--330,
  1999.

\bibitem[Shang et~al.(2024{\natexlab{a}})Shang, Wang, and Stern]{shang2024two}
Mingfeng Shang, Shian Wang, and Raphael Stern.
\newblock A two-condition continuous asymmetric car-following model for
  adaptive cruise control vehicles.
\newblock \emph{IEEE Transactions on Intelligent Vehicles}, 2024{\natexlab{a}}.

\bibitem[Aguilar and Wang(2024)]{aguilar2024energy}
Jose~Acedo Aguilar and Shian Wang.
\newblock Energy impacts of traffic-smoothing cruise controllers on mixed
  traffic.
\newblock In \emph{2024 Forum for Innovative Sustainable Transportation Systems
  (FISTS)}, pages 1--6. IEEE, 2024.

\bibitem[Jin and Orosz(2014)]{jin2014dynamics}
I~Ge Jin and G{\'a}bor Orosz.
\newblock Dynamics of connected vehicle systems with delayed acceleration
  feedback.
\newblock \emph{Transportation Research Part C: Emerging Technologies},
  46:\penalty0 46--64, 2014.

\bibitem[De~Souza and Stern(2021)]{de2021calibrating}
Felipe De~Souza and Raphael Stern.
\newblock Calibrating microscopic car-following models for adaptive cruise
  control vehicles: Multiobjective approach.
\newblock \emph{Journal of Transportation Engineering, Part A: Systems},
  147\penalty0 (1):\penalty0 04020150, 2021.

\bibitem[Ahn et~al.(2002)Ahn, Rakha, Trani, and Van~Aerde]{ahn2002estimating}
Kyoungho Ahn, Hesham Rakha, Antonio Trani, and Michel Van~Aerde.
\newblock Estimating vehicle fuel consumption and emissions based on
  instantaneous speed and acceleration levels.
\newblock \emph{Journal of Transportation Engineering}, 128\penalty0
  (2):\penalty0 182--190, 2002.

\bibitem[Sun et~al.(2022)Sun, Wang, Shao, Sun, and Levin]{sun2022energy}
Wenbo Sun, Shian Wang, Yunli Shao, Zongxuan Sun, and Michael~W Levin.
\newblock Energy and mobility impacts of connected autonomous vehicles with
  co-optimization of speed and powertrain on mixed vehicle platoons.
\newblock \emph{Transportation Research Part C: Emerging Technologies},
  142:\penalty0 103764, 2022.

\bibitem[Shang et~al.(2024{\natexlab{b}})Shang, Wang, Li, and
  Stern]{shang2024interaction}
Mingfeng Shang, Shian Wang, Tianyi Li, and Raphael Stern.
\newblock Interaction-aware model predictive control for automated vehicles in
  mixed-autonomy traffic.
\newblock In \emph{2024 IEEE Intelligent Vehicles Symposium (IV)}, pages
  317--322. IEEE, 2024{\natexlab{b}}.

\end{thebibliography}

\end{document}